\documentclass[twocolumn,trackchanges]{aastex631}
\usepackage{textcomp}
\usepackage{amsmath}
\usepackage{graphicx}

\usepackage{hyperref}

\begin{document}

%\title{Obliquity of Black Hole Magnetosphere and its Impact on Accretion Dynamics}
%\title{Magnetic Field Deformations and Their Impact on Black Hole Accretion Dynamics}
\title{Equatorially Asymmetric Magnetic Fields and Their Impact on Black Hole Accretion Dynamics}
%Axisymmetric Magnetic Field Deformations and Their Impact on Black Hole Accretion Dynamics
%Effects of Equatorially Asymmetric Magnetic Fields on Relativistic Accretion Flows
%Accretion Dynamics Around a Kerr Black Hole with Axisymmetrically Deformed Magnetic Fields
%Influence of Axisymmetric Magnetic Perturbations on Black Hole Accretion and Outflows
%Accretion Flow Response to Non-Uniform but Axisymmetric Magnetic Field Configurations
%Exploring Non-Uniform Axisymmetric Magnetic Structures in Black Hole Accretion Flows

\author[0000-0002-7827-4517]{Ishika Palit}
\affiliation{Institute of Astronomy and Department of Physics, National Tsing Hua University, Hsinchu 30013, Taiwan}
%\email{ishi2694@gmail.com}

\author[0000-0002-4064-0446]{Indu K. Dihingia}
\affiliation{Institute of Fundamental Physics and Quantum Technology, \& School of Physical Science and Technology, Ningbo University, Ningbo, Zhejiang 315211, China}
\affiliation{Tsung-Dao Lee Institute, Shanghai Jiao-Tong University, 1 Lisuo Road, Shanghai, 201210, People’s Republic of China}

\author[0000-0002-8131-6730]{Yosuke Mizuno}
\affiliation{Tsung-Dao Lee Institute, Shanghai Jiao-Tong University, 1 Lisuo Road, Shanghai, 201210, People’s Republic of China}
\affiliation{School of Physics \& Astronomy, Shanghai Jiao-Tong University, Shanghai, 800 Dongchuan Road, 200240, People's Republic of China}
\affiliation{Key Laboratory for Particle Physics, Astrophysics and Cosmology (MOE), Shanghai Key Laboratory for Particle Physics and Cosmology, Shanghai Jiao-Tong University, 800 Dongchuan Road, Shanghai, 200240, China}

\author[0000-0003-3269-4660]{Hsiang-Yi Karen Yang}
\affiliation{Institute of Astronomy and Department of Physics, National Tsing Hua University, Hsinchu 30013, Taiwan}
\affiliation{Physics Division, National Center for Theoretical Sciences, Taipei 10617, Taiwan}

\begin{abstract}
We investigate the impact of equatorial asymmetry in the magnetic field geometry on accretion dynamics around a spinning black hole using axisymmetric general relativistic magnetohydrodynamic simulations. We consider a Fishbone--Moncrief torus orbiting a Kerr black hole with spin parameter $a = 0.9375$, threaded by large-scale magnetic fields that are asymmetric about the equatorial plane. The degree of equatorial asymmetry in the magnetic field is parametrized by an angle, with values of $30^\circ$, $45^\circ$, and $60^\circ$. We examine how this equatorially asymmetric initial magnetic field configuration influences the magnetic field structure, accretion flow morphology, and angular momentum transport across a range of initial plasma beta values ($\beta = 0.007, 0.005, 0.001$). We find that such deformation in the magnetic field leads to noticeable changes in the inner disk structure, asymmetric outflow patterns in the poloidal plane, and time-dependent variations in accretion rates. These effects are generally more pronounced at lower beta values, where magnetic pressure dominates; in particular, the $30^\circ$ case at $\beta = 0.001$ exhibits strong and persistent asymmetric inflows and outflows. Our results demonstrate that equatorially asymmetric magnetic field configurations can significantly influence the structure and variability of relativistic accretion flows.  These findings motivate future extensions to full three-dimensional studies, where black hole magnetosphere can be explored in a more general setting.
\end{abstract}


\section{Introduction}
Black holes (BHs) in astrophysical environments are often expected to be misaligned with the surrounding magnetic field or accretion flow due to complex formation histories and environmental dynamics \citep[e.g.,][]{2013Sci...339...49M,1975ApJ...195L..65B,2006MNRAS.368.1196L}. This is particularly true in systems such as Active Galactic Nuclei (AGN) and X-ray binaries, where misalignments can arise from galaxy mergers, black hole mergers, or fallback of chaotic material \citep[e.g.,][]{2019MNRAS.487.3488B,2008MNRAS.391L..15M,2015ApJ...800...17F}. Modeling such oblique configurations is therefore essential for building realistic astrophysical scenarios.

In accreting systems, the evolution of black hole magnetospheres, especially in magnetically arrested disk (MAD) states, plays a crucial role in governing the dynamics of relativistic jets and the efficiency of energy extraction \citep{2011MNRAS.418L..79T,2003PASJ...55L..69N}. Unlike neutron stars, black holes can expel magnetic flux through reconnection processes near the event horizon, potentially driving powerful flares in the X-ray and gamma-ray bands \citep{2012MNRAS.419..573M,2006MNRAS.368.1561M,2012MNRAS.423.3083M,1997ApJ...489..865E,2008ApJ...688..555G,2022MNRAS.517.5032D,2024A&A...688A..82J,2025arXiv250712789J}. The orientation of the magnetic field relative to the black hole spin axis has been shown to affect jet launching, collimation, and stability, with implications for observed jet morphologies and variability \citep{2004ApJ...615..389M,2019MNRAS.487..550L,2021MNRAS.505.3596D,2024A&A...687A.185J,2022Univ....8...85M,2025arXiv250713680J}.

A few recent general relativistic magnetohydrodynamic (GRMHD) studies have explored how oblique magnetic fields alter magnetospheric structures. For example, \citet{2024ApJ...968L..10S} investigated the evolution of a misaligned split-monopole field around a non-accreting spinning black hole. They demonstrated that the inclined current sheet tends to realign with the equatorial plane over time, with an exponential decay of the inclination angle dependent on the square of the black hole's spin. Similarly, \citet{2021AN....342..357K} studied the dynamics of charged particles in large-scale inclined magnetic fields near a rotating black hole, showing that ionization-induced destabilization can launch relativistic outflows along jet-like trajectories, with increased inclination enhancing particle acceleration and the fraction of unbound orbits.

Despite these insights, most existing work has focused on idealized or non-accreting systems. The impact of magnetic field obliquity on accreting black holes, especially in high-magnetization regimes, remains relatively unexplored. In particular, how inclined magnetic fields interact with turbulent plasma in GRMHD simulations, and how they affect accretion rates, field topology, and energy extraction, warrants further study.

In MAD states, the strong magnetic flux near the black hole tends to realign or suppress inclined fields through reconnection and flux eruptions, making it difficult to sustain long-term obliquity \citep[see][and references therein]{2025ApJS..277...16D,2011MNRAS.418L..79T}.
Additionally, achieving and maintaining the MAD state requires fine-tuned conditions, such as a continuous supply of large-scale poloidal flux, which may not be common in all astrophysical environments. In contrast, SANE disks allow inclined fields to persist and interact with turbulent flows, making them more suitable for systematically studying the effects of magnetic obliquity on accretion and jet dynamics. Even in the absence of a fully 3D oblique field, equatorial asymmetric or deformed magnetic fields can influence the accretion dynamics and jet formation. Modeling such equatorial deformations is therefore a useful first step in studying asymmetric magnetospheres in GRMHD simulations. Such deformations in the magnetic field can create asymmetric current sheets and flux imbalances near the horizon, which influence jet launching, collimation, and variability even in 2D simulations \citep{2024A&A...687A.185J, 2020bhns.work..107K, 2021MNRAS.504.6076R}.

Persistent asymmetric jets or outflows are of interest because several observed systems, such as radio galaxies, X-ray binaries, and Tidal Disruption Events, exhibit one-sided jets, bent morphologies, or asymmetries in brightness and structure that cannot be explained by Doppler boosting alone. Such features may arise from intrinsic asymmetries in the magnetosphere or accretion flow, possibly driven by inclined magnetic fields. Studying these configurations helps us connect simulation results to observed jet variability, wiggling, or asymmetry in sources like M87, SS~433, and certain FR I/II radio galaxies \citep{2015ApJ...811...92C,2020MNRAS.499..362C,2023Natur.621..711C,2001ApJ...553..955F}.

In this work, we perform a suite of axisymmetric (2D, two-dimensional) GRMHD simulations of a geometrically thick torus accreting onto a spinning Kerr black hole with an imposed equatorial asymmetry in the initial magnetic field configuration. We explore a range of initial plasma beta values and deformation angles to assess how departures from equatorial magnetic field symmetry influence accretion flow morphology, magnetic field evolution, angular momentum transport, and jet behavior. Our results provide new insights into the dynamic role of equatorially asymmetric magnetic fields in shaping black hole accretion systems and offer potential observational implications for interpreting jet variability and asymmetry.

The paper is organized as follows. In Section~\ref{sec-2}, we describe the numerical methods and simulation setup, including details of the equatorially asymmetric vector potential and magnetic field initialization (Section~\ref{sec-2.2}), as well as our approach to classifying outflows (Section~\ref{sec-2.3}). Section~\ref{sec-3} presents the main results: starting with the initial configuration (Section~\ref{sec-3.1}), we analyze the time-averaged disk and jet structures (Section~\ref{sec-3.2}), and systematically explore the effects of the equatorial asymmetry angle (Section~\ref{sec-3.3}) and magnetic field strength (Section~\ref{sec-3.4}), followed by a discussion of mass accretion rates (Section~\ref{sec-3.5}). We summarize our key findings in Section~\ref{sec-4} and discuss their astrophysical implications in Section~\ref{sec-5}. Additional technical details and supporting analyses are provided in the Appendices.

\section{Numerical Simulation}
\label{sec-2}

\subsection{\MakeUppercase{ Simulation Setup}}
\label{sec-2.1}
\noindent
In this study, we simulate black hole accretion flows using a GRMHD framework, assuming a fixed Kerr spacetime with dimensionless spin parameter $a = 0.9375$. The black hole mass $M$, gravitational constant $G$, and speed of light $c$ define the gravitational radius $r_g = GM/c^2$, and the event horizon is located at $r_H = (1 + \sqrt{1 - a^2})\, r_g$. We employ the ideal \texttt{BHAC} code \citep{2017ComAC...4....1P,2019A&A...629A..61O} to solve the GRMHD equations in modified Kerr–Schild (MKS) coordinates, which is a covariant, horizon-penetrating spherical coordinate system commonly used in GRMHD simulations. \\

The computational domain spans a radial range of \( r = [0.9\,r_H, 2500\,r_g]\), and in the polar direction, the grid covers the angular range \( \theta \in [0^\circ,\, 180^\circ] \). Grid resolution is $1024 \times 512$ cells, logarithmically spaced in the radial direction and uniformly spaced in angle. To ensure numerical stability and avoid coordinate singularities at the polar axis (\( \theta = 0 \) and \( \theta = \pi \)), the angular domain is restricted to exclude the immediate vicinity of the poles. This angular restriction enhances numerical stability and is suitable for studying systems with equatorially deformed magnetic fields.

\noindent
To avoid the coordinate singularity at the polar axis, a small conical region around $\theta = 0$ and $\theta = \pi$ is excised. This implements a soft polar boundary condition, similar to that described in \citet{2012ApJ...744..187S}, which behaves approximately like a reflective wall while allowing minimal magnetic flux leakage through the excised polar region. The amount of flux passing through this narrow excision zone is negligible compared to the total horizon-threading flux, and therefore the evolution of the disk and the deformed magnetosphere is not significantly affected by the boundary treatment. The magnetic field is almost divergence-free everywhere the equations are solved, and no numerical or physical violation of $\nabla.B$ occurs within the evolved region. This ensures that the time-dependent horizon flux and asymmetric outflows we report reflect the physical dynamics of the system rather than artifacts of the polar axis treatment.

\noindent 
%Boundary conditions are chosen to enforce reflection symmetry at the $\theta$ boundaries and prevent unphysical inflows.
The inner and outer radial boundaries are set to outflow conditions, allowing jets and outflows to leave the domain freely without bouncing back during the simulation.
The plasma beta, $\beta$, is defined as the ratio of $P_{\rm gas}/P_{\rm mag} =  2P / B^2$, where $P$ is the gas pressure and $B$ is the magnetic field strength.
 
In this work, we focus on magnetized accretion flows with low plasma-$\beta$. The initial plasma-$\beta$ is set to - ($\beta = 0.001, 0.005, 0.007$). Our aim is to investigate magnetically driven asymmetric winds, which are most relevant in regions where magnetic pressure becomes dynamically important. Such low-$\beta$ states naturally develop in the inner disk and in magnetically arrested configurations. The magnetic field is normalized using the global maximum gas pressure and magnetic energy density, consistent with the HARM scheme \citep{2003ApJ...589..444G}, ensuring that the target plasma-$\beta$ is correctly imposed throughout the domain. The initial plasma-$\beta$ is low, but around the torus it quickly grows to $\beta \sim 10^{2}$, and continues increasing outward, reaching values of $\beta \sim 10^{15}$ near the outer grid radius ($r \sim 2500\,r_g$), following an approximately logarithmic trend.\\

\noindent The accretion flow is initialized with a relativistic Fishbone--Moncrief torus in hydrostatic equilibrium \citep{1976ApJ...207..962F,2024ApJ...970..172U}, centered around a spinning black hole with a dimensionless spin parameter \( a = 0.9375 \). The torus extends from an inner edge at \( r_{\mathrm{in}} = 4 \) to a pressure maximum located at \( r_{\mathrm{max}} = 12 \). The fluid is described by an ideal equation of state with adiabatic index \( \gamma = 4/3 \), and the initial entropy parameter is set to \( \kappa = 10^{-3} \), which is later renormalized to ensure the desired peak density and pressure.

\noindent The simulations are carried out in a geometrized unit system. 
In this system, length and time are expressed in units of gravitational radius \( r_g = G M_{\mathrm{BH}}/c^2 \) and gravitational time \( t_g = G M_{\mathrm{BH}}/c^3 \).

\begin{figure*}[ht]
    \centering
    \includegraphics[width=\linewidth]{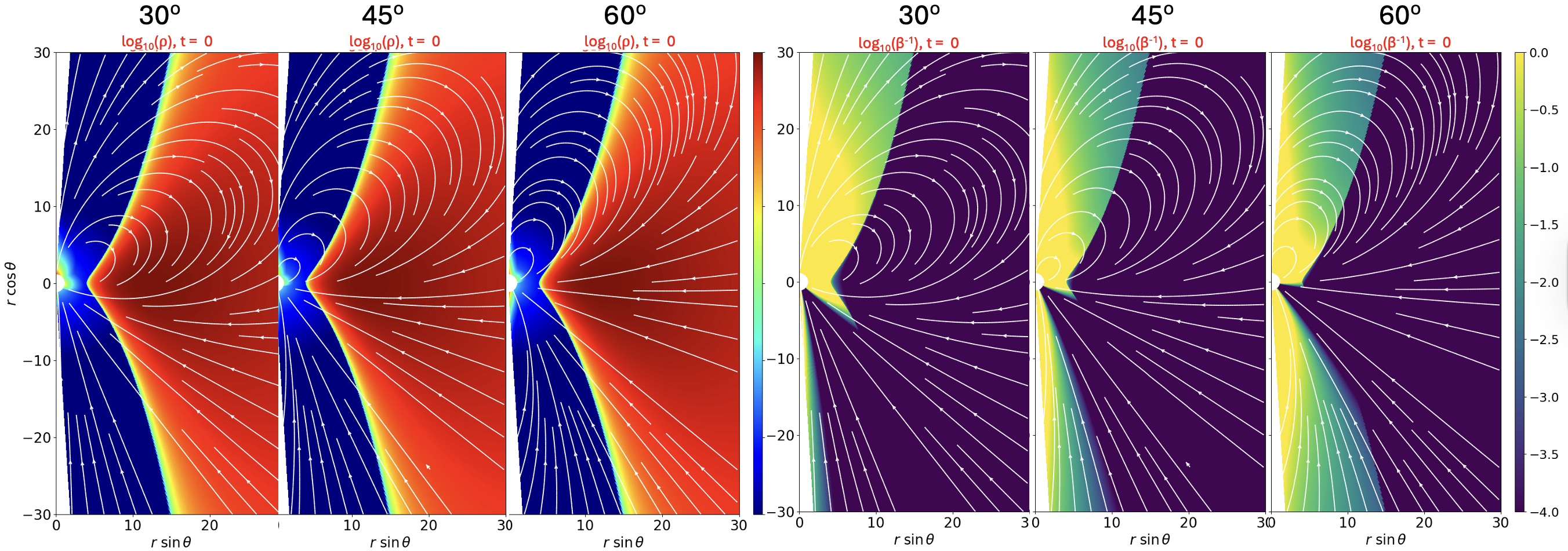}
    \caption{Initial magnetic field configuration for simulations with deformation angles \( 30^\circ,\, 45^\circ,\, \text{and } 60^\circ \), shown at \( t = 0 \) for \( \beta = 0.005 \). The left three panels show the logarithmic rest-mass density distributions in the poloidal plane for each deformation angle, while the right three panels display the corresponding plasma-$\beta = 2P/B^{2}$ parameter. White lines overlaid on each panel represent magnetic field streamlines, highlighting the initial magnetic topology.}
    \label{fig:figure_1}
\end{figure*}

\subsection{\MakeUppercase {Equatorial Asymmetric Magnetic Field}}
\label{sec-2.2}

\noindent The equatorially asymmetric magnetic vector potential plays a significant role in shaping the dynamics of the accretion flow. By introducing a polar offset \( z \) in the angular dependence of the vector potential \( A(\mathbf{x}) \), we break the equatorial symmetry and localize the magnetic structure around a chosen deformation angle, allowing us to explore the impact of such magnetic fields in moderately magnetized (SANE) accretion states.

\noindent Such configurations mimic realistic astrophysical environments, where magnetic flux may be advected from misaligned accretion flows or accumulated with a net deformation. By varying \( z \), we investigate how different magnetic geometries influence the efficiency of magnetic flux accumulation, jet launching, and the symmetry of outflows.

\begin{table}
\centering
%\begin{flushleft}
{\large
\begin{tabular}{|c|c|c|c|}
\hline
\textbf{deformation} & \textbf{0.001} & \textbf{0.005} & \textbf{0.007} \\
\hline
30° & Test 1 & Test 4 & Test 7 \\
45° & Test 2 & Test 5 & Test 8 \\
60° & Test 3 & Test 6 & Test 9 \\
\hline
\end{tabular}
}
\caption{Grid of simulation test runs showing different combinations of magnetic field deformation angle and initial plasma-$\beta$ values. }

\label{table:grid_tests_all_beta}
%\end{flushleft}
\end{table}

\noindent We initialize the magnetic field by prescribing a toroidal vector potential \( A_\phi(\mathbf{x}) \), localized around a specific polar deformation angle \( z_{o} \). This breaks equatorial symmetry and allows controlled deformation of the magnetic structure. The expression used is:

\begin{equation}
   A_\phi(\mathbf{x}) = f_0 \, \exp\left(-\frac{1}{\sin^2(\theta + z_{o})}\right) \cdot \frac{1 - \cos^2(\theta + z_{o})}{r \sin(\theta + z_{o})}, 
\end{equation}

\noindent where \( \theta \) is the polar angle, \( r \) is the radial coordinate, and \( f_0 \) is a normalization factor set by the initial plasma \( \beta \). The angle of deformation of the magnetic field with respect to the polar axis is varied across the simulations as $z_{o} = 30^\circ,\, 45^\circ,\, \text{and } 60^\circ$, allowing us to systematically study how the degree of deformation affect the magnetic field geometry and the resulting flow dynamics. The exponential term in the definition of \( A_\phi(\mathbf{x}) \) ensures that the magnetic field is tightly confined to a narrow angular range around \( \theta + z \), minimizing artificial field structures near the poles and concentrating magnetic pressure in targeted regions.

Table~\ref{table:grid_tests_all_beta} summarizes the full set of simulation runs as a function of deformation angle and initial magnetization. Our fiducial run corresponds to \texttt{Test 5} in the Table \ref{table:grid_tests_all_beta}, with deformation angle \( 45^\circ \) and \( \beta = 0.005 \). All our simulation models were evolved up to \( t = 12{,}000\,t_g \).

\begin{figure}[ht]
    \centering
    \includegraphics[width=\linewidth]{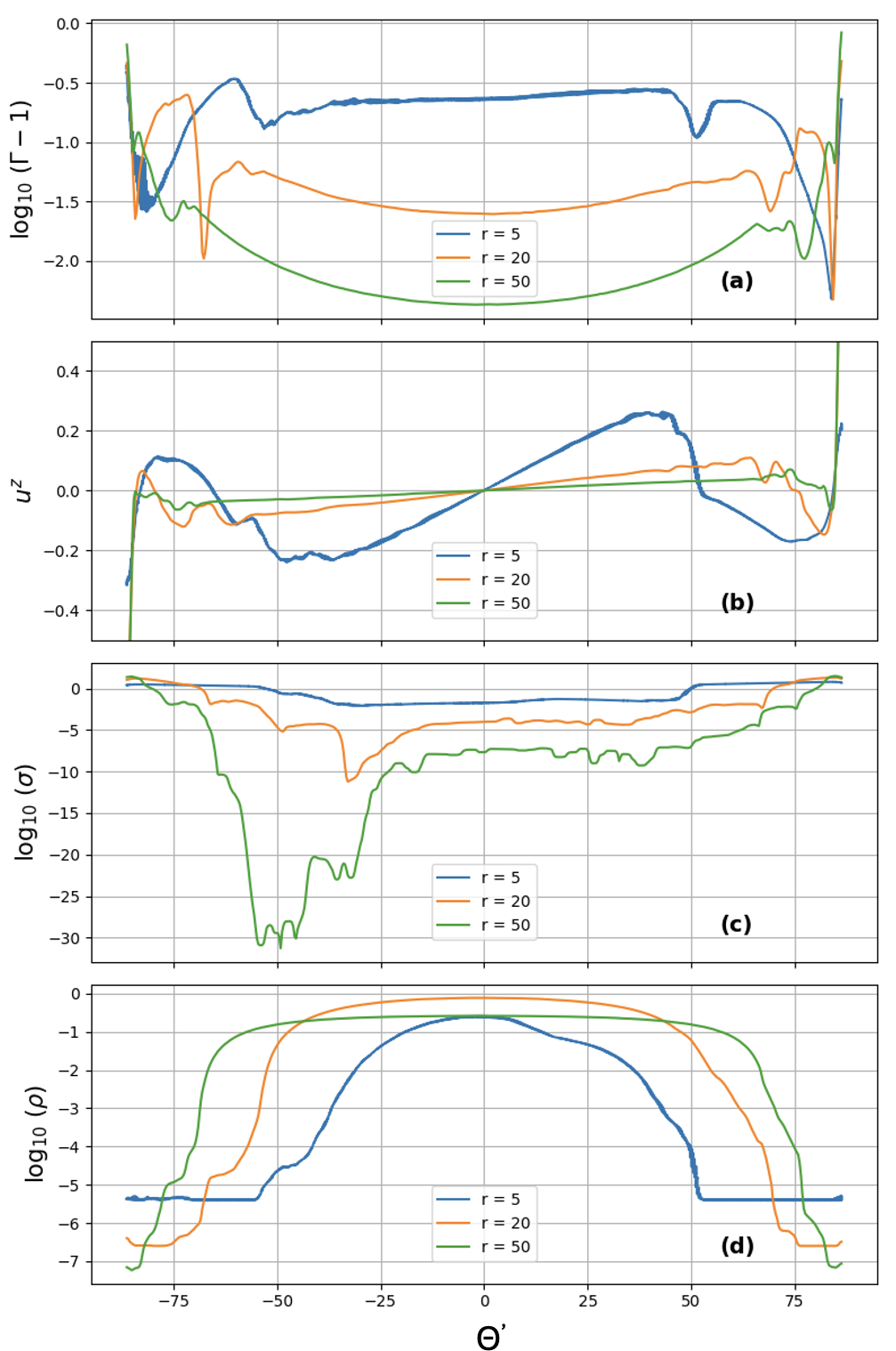}
    \caption{
    Angular profiles of time-averaged quantities at fixed radii for the simulation with initial $\beta = 0.005$ and deformation angle $45^\circ$. Plotted are the variations of Lorentz factor, velocity, magnetization parameter, and density as functions of $\theta'$ at three representative radii ($r = 5$, $20$, and $50$). The values are averaged over the quasi-steady period $t = 9,000$ to $11,000$ $t_{g}$.}
    \label{fig:figure_2}
\end{figure}
%Negative or nonphysical values have been excluded by applying a floor before taking the logarithm.

\subsection{\MakeUppercase {Classification of Jet and Wind}}
\label{sec-2.3}

In GRMHD simulations, jets and winds emerge as distinct outflow components. Jets are narrow, relativistic, and magnetically dominated (\( \sigma \gg 1 \)), launched along the rotation axis and often powered by the Blandford–Znajek mechanism \citep{1977MNRAS.179..433B}. Winds, on the other hand, are slower, broader, and originate from the disk corona or funnel wall, driven by a combination of magnetic, thermal, and centrifugal forces \citep{2007ApJ...662..835M}.\\

Although the general morphology of black hole accretion systems is qualitatively understood, a precise and universally accepted quantitative definition of the jet/wind remains elusive. Several criteria have been proposed in the literature, including a purely geometric Bernoulli parameter \citep{2004ApJ...611..977M}, fluid-based definitions that include thermal and kinetic contributions \citep{2012MNRAS.426.3241N,2016A&A...586A..38M}, and magnetohydrodynamic extensions that also incorporate magnetic energy in the total budget \citep{2010MNRAS.408..752P,2015ApJ...804..101Y}.

In our simulations, we distinguish between the jet and wind based on physical criteria derived from local fluid and electromagnetic properties. This classification enables separate diagnostics for mass, angular momentum, and energy fluxes originating from different zones of the accretion flow.
A common quantitative measure for identifying energetically unbound outflows is the fluid Bernoulli parameter (\cite{2012MNRAS.426.3241N}):
\begin{equation}
    \mu = \frac{-T^r_t}{\rho u^r},
\end{equation}
where \( T^r_t \) is the radial component of the energy-momentum tensor, \( \rho \) is the rest-mass density, and \( u^r \) is the radial component of the fluid four-velocity. If \( \mu > 1 \), the fluid is energetically unbound and can escape to infinity. \\
In our simulations, zones are classified into jet and wind regions based on physical criteria involving magnetization and energy flux \citep{2020MNRAS.495.1549N,2021MNRAS.505.3596D}.

\begin{itemize}
    \item Jet Region: A zone is considered part of the jet if it satisfies high magnetization:
        \begin{equation}
            \sigma = \frac{b^2}{\rho} > 1,
        \end{equation}
        where \( b^2 \) is the magnetic energy density and \( \rho \) is the rest-mass density. This condition selects regions where electromagnetic energy dominates, typical of strongly magnetized jet interiors.

    Energetically unbound, relativistic outflows usually have $\mu  > \mu_{\rm thresh}$. The default threshold is set in our simulations as: $\mu_{\rm thresh} = 2.0$. This corresponds approximately to a Lorentz factor \( \Gamma > 2 \), identifying a relativistically outflowing medium.

    \item Wind Region: Zones not classified as part of the jet may be identified as part of the wind if they satisfy the unbound enthalpy condition:
    \begin{equation}
        -h u_t > h_{\rm thresh},
    \end{equation}
    where \( h \) is the total specific enthalpy and \( u_t \) is the covariant time component of the four-velocity. The threshold is set as \( h_{\rm thresh} = 1.0 \), selecting regions of the medium with sufficient thermal or magnetic energy to overcome gravitational binding.
\end{itemize}

These combined magnetization and energy-based criteria provide a physically motivated and numerically robust method to distinguish between jet and wind regions in GRMHD simulations. The selected regions correspond to either highly magnetized or energetically unbound relativistic outflows, typically associated with polar jets (e.g., \citealt{2021ApJ...914...55W,2020ARA&A..58..407D,2013Sci...339...49M,2024ApJ...972...18A}).

\smallskip
\noindent
Fluxes such as the mass accretion rate and angular momentum are computed by integrating over each region, weighted by the appropriate fluid quantities and radial Jacobian factors. Outflows are defined as regions where the radial four-velocity satisfies \( u^r > 0 \), while inflows are characterized by \( u^r < 0 \).

%\newpage
\section{Results}
\label{sec-3}

\subsection{\MakeUppercase {Initial configuration}}
\label{sec-3.1}

Figure~\ref{fig:figure_1} illustrates the initial configuration for our fiducial model with \( \beta = 0.005 \), shown across three magnetic field deformation angles. The first three out of six panels display the rest-mass density in the poloidal plane, showing a dense torus near the equator and lower-density regions along the polar axis. The rest of the panels show the inverse of plasma-$\beta$, which is highest along the polar regions and decreases toward the equatorial torus. White magnetic field lines are overplotted to highlight the inclined field geometry, which seeds asymmetry in the evolving flow.

This figure highlights how the initial magnetic topology varies with deformation, which plays a crucial role in shaping the subsequent dynamics of the accretion flow and outflows. The equatorially deformed magnetic fields are expected to introduce asymmetries in the $z$-direction and corona, potentially influencing mass and energy outflows. Furthermore, the initial spatial distribution of \( \beta \) provides insight into regions of strong magnetic dominance, which are key to understanding the onset of magnetically driven acceleration and collimation mechanisms.

\subsection{\MakeUppercase {Time-averaged disk structure}}
\label{sec-3.2}

To study the underlying flow structure, we performed time averaging over various intervals. In this paper, we present results based on the later time interval, 9000$–$11{,}000 $t_{g}$, to best capture the quasi-steady state of the system.

Figure~\ref{fig:figure_2} shows time-averaged 1D radial profiles of key physical quantities for the fiducial case with \( \beta_{0} = 0.005 \) and magnetic field deformation angle \( 45^\circ \). The time averaging, performed over $ t = 9000 - 11{,}000\,t_{g}$, effectively smooths out transient fluctuations and highlights persistent features in the flow.

\noindent  Panels (a)–(d) of Figure~\ref{fig:figure_2} display the time-averaged Lorentz factor \( \log_{10}(\Gamma - 1) \), vertical velocity \( u^{z} \), magnetization \( \log_{10}(\sigma) \), and rest-mass density \( \log_{10}(\rho) \), respectively, measured at fixed radii \( r = 5\, r_g \), \( 20\, r_g \), and \( 50\, r_g \). The profiles are shown as functions of $\theta' \in [-90^\circ, 90^\circ]$, where $\theta' = \theta - 90^\circ$ and $\theta$ is measured from the polar axis in the simulation. This convention places the equatorial plane at $\theta' = 0^\circ$ and the two poles at $\theta' = \pm 90^\circ$, providing a symmetric view of the disk and outflow structure in the $x$--$z$ plane and tracing the distribution of relativistic kinetic energy.

Panel (a) of Figure~\ref{fig:figure_2} shows the 1D profile of Lorentz factor ($\Gamma$). At \( r = 5\,r_g \), a gradual increase in \( \log_{10}(\Gamma - 1) \) is observed from the disk midplane up to \( \sim 50^\circ \), indicating progressive acceleration toward the polar region. This trend, evident in the time-averaged profiles over $ t = 9000 - 11{,}000 \,  t_{g }$, reflects the presence of moderately relativistic, wind-like outflows emerging from the inner disk.

Panel (b) of Figure~\ref{fig:figure_2} shows the vertical velocity component \( u^{z} \) in the \( \phi = 0 \) plane (i.e., the \( x\text{--}z \) or poloidal plane). The velocity is positive in the upper hemisphere (\( \theta' = 0^\circ\text{--}90^\circ \)) and negative in the lower hemisphere (\( \theta' = -90^\circ\text{--}0^\circ \)), consistent with a bipolar outflow directed along both \( +z \) and \( -z \) axes ($\sim 0.2c$). The magnitude of \( u^{z} \) is highest at \( r = 5\,r_g \), and gradually decreases at larger radii (\( r = 20\,r_g \) and \( 50\,r_g \)), suggesting that the vertical acceleration is strongest near the base of the outflow.

Panel (c) shows the magnetization \( \log_{10}(\sigma) \) as a function of $\theta'$. A clear dip in magnetization is observed in the lower hemisphere (\( \theta' = -90^\circ\text{--}0^\circ \)) at \( r = 20\,r_g \) and becomes more pronounced at \( r = 50\,r_g \). In contrast, \( \sigma \) remains relatively flat at \( r = 5\,r_g \), indicating that magnetic energy is more symmetrically distributed closer to the black hole, while asymmetries develop further out.

Panel (d) shows the density profile, which exhibits a characteristic torus-like shape at \( r = 5\,r_g \), concentrated around the equatorial plane. However, at larger radii (\( r = 20\,r_g \) and \( 50\,r_g \)), the profile becomes progressively flatter. This flattening reflects the radial expansion and vertical broadening of the disk, along with the onset of mass-loading into outflows at higher latitudes.

\begin{figure*}[ht]
    \centering
    \includegraphics[width=\linewidth]{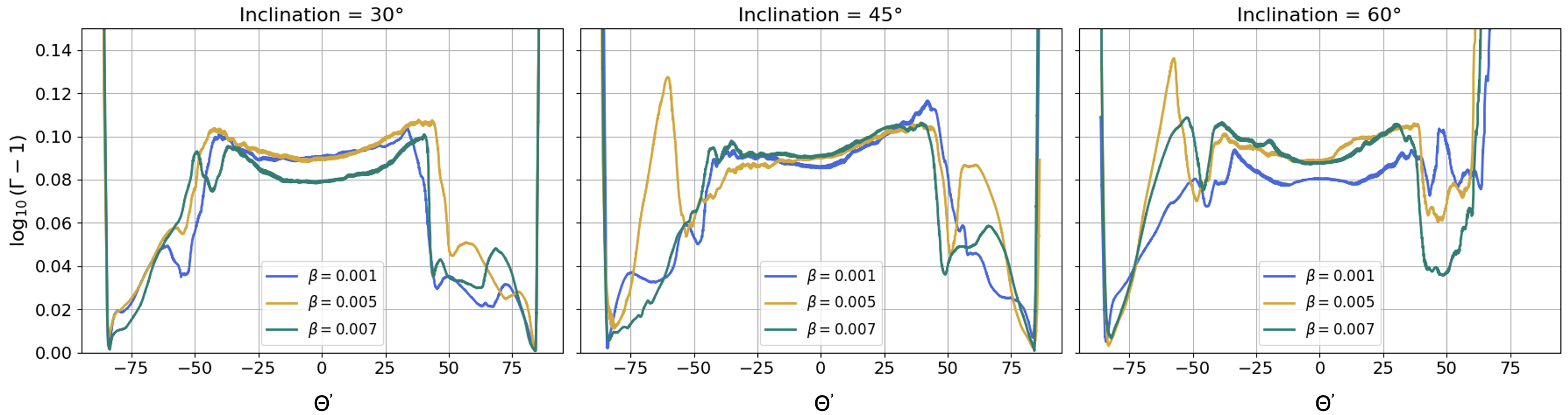}
    \caption{Time-averaged Lorentz factor profile, \( \log_{10}(\Gamma - 1) \) as a function of \( \theta' \) at \( r = 5\,r_g \). It shows the variation with plasma beta (\( \beta = 0.001,\, 0.005,\, 0.007 \)) for deformation angles \( 30^\circ \), \( 45^\circ \), and \( 60^\circ \), respectively. All profiles are time-averaged over \( t = 9000\text{--}11{,}000\,t_g \).}
    \label{fig:figure_3}
\end{figure*}

%---------------------------------------------------------
\subsection{\MakeUppercase {Variation with deformation}}
\label{sec-3.3}

Magnetic field geometry plays a crucial role in shaping the dynamics of accretion flows and associated outflows. In particular, the deformation of the magnetic field lines relative to the rotation axis can significantly impact the structure, strength, and symmetry of the resulting jets and winds. In this section, we show how varying the magnetic field deformation influences key flow characteristics, focusing on the Lorentz factor and the morphology of 2D flow structures.

Figure~\ref{fig:figure_3}  illustrates the dependence of Lorentz factor ($\Gamma$) on plasma beta for deformation angles: \( 30^\circ \), \( 45^\circ \), and \( 60^\circ \) respectively.

We examine $\log_{10}(\Gamma - 1)$ as a function of  $\theta' \in [-90^\circ, 90^\circ]$ to trace the distribution of relativistic kinetic energy in the flow. 
Since $(\Gamma - 1)$ represents the specific kinetic energy, this diagnostic highlights energetic outflows. In our case, $\log_{10}(\Gamma - 1)$ increases sharply toward the poles and dips significantly between $\theta' \approx \pm45^\circ$ to $\pm75^\circ$, indicating a low-energy sheath surrounding a relativistic spine. This sharp angular variation suggests a collimated structure rather than an isotropic distribution. The stronger magnetic tension facilitates more efficient collimation and acceleration of the jet spine, leading to higher Lorentz factors near the poles. In the following sections, we further discuss how different variables are influenced by varying deformation and plasma-$\beta$. 

\begin{figure*}[ht]
    \centering
    \textbf{deformation angle, $30^{\circ}$} \\
    \includegraphics[width=\linewidth]{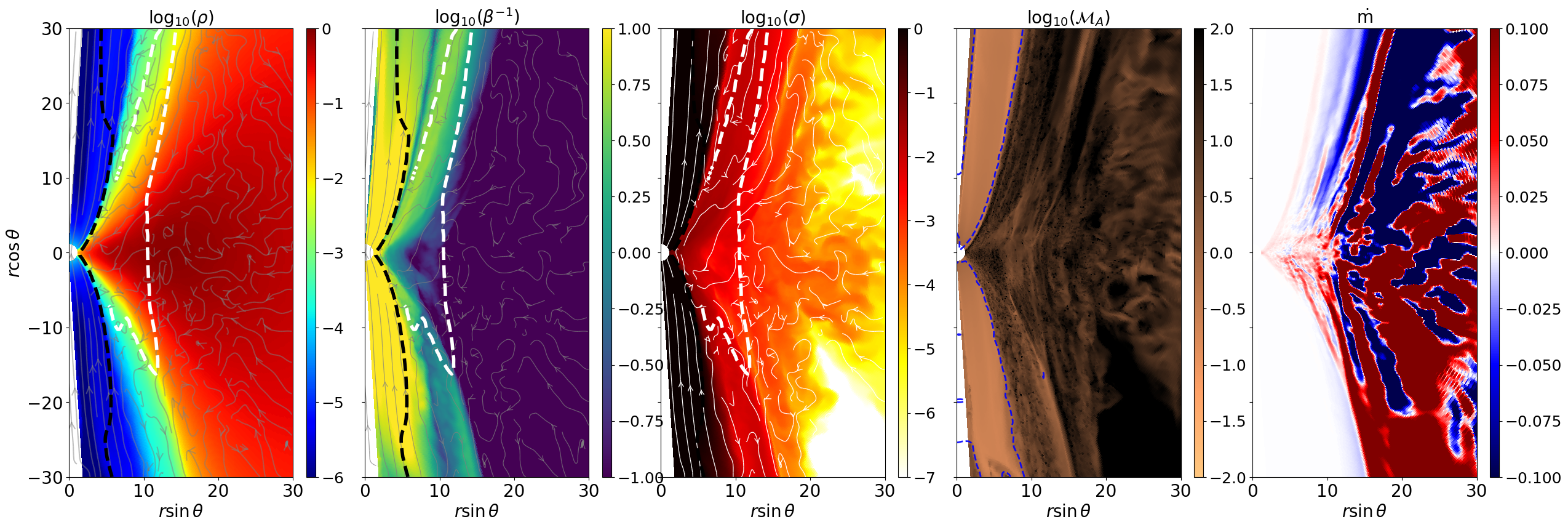}
    
    \textbf{deformation angle, $45^{\circ}$} \\
    \includegraphics[width=\linewidth]{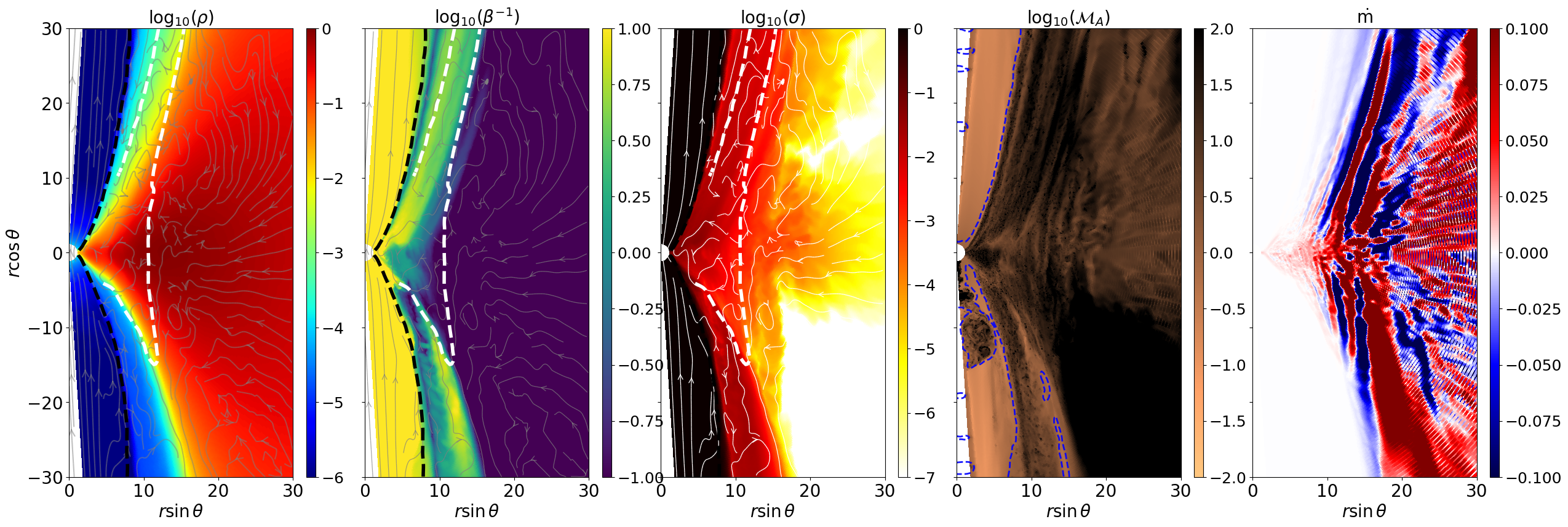}

    \textbf{deformation angle, $60^{\circ}$} \\
    \includegraphics[width= \linewidth]{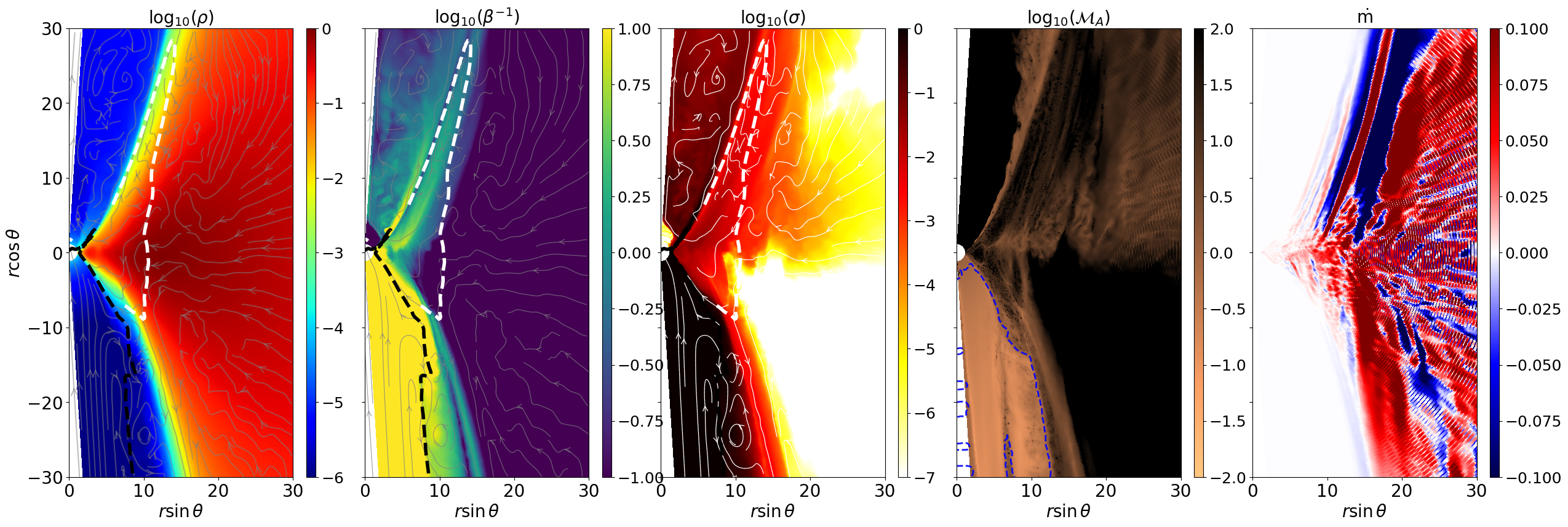}
    \caption{Time-averaged 2D distribution of key physical quantities for the fiducial run with $\beta = 0.005$, averaged over $t = 9000$–$11000$ $t_{g}$ across the simulation domain (\( x\text{--}z \) plane). From left to right: rest-mass density $\rho$, plasma-$\beta^{-1}$, magnetization parameter $\sigma = b^2/\rho$, Alfv\'{e}n Mach number $\mathcal M_{\rm A}$, and mass flux, $\dot{\rm{m}}$ = -$\sqrt{-g}\rho u^{r}$ . Each row corresponds to a different deformation: (top) $30^\circ$, (middle) $45^\circ$, and (bottom) $60^\circ$. The first three panels in each row are overplotted with magnetic field contours, as well as the jet boundary (black dashed line) and wind boundary (white dashed line).
    }
    \label{fig:figure_4}
\end{figure*}

\begin{figure*}[ht]
    \centering

    \textbf{plasma $\beta_{0} = 0.001$} \\
    \includegraphics[width=\linewidth]{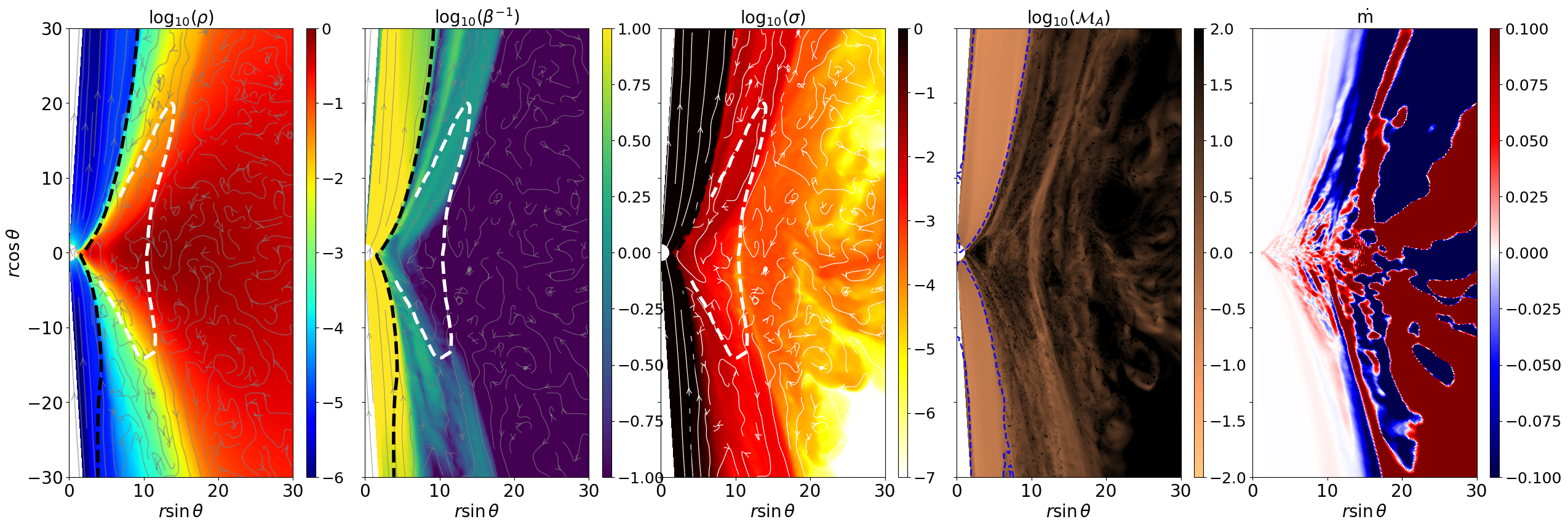} \\[1ex]

    \textbf{plasma $\beta_{0} = 0.005$} \\
    \includegraphics[width=\linewidth]{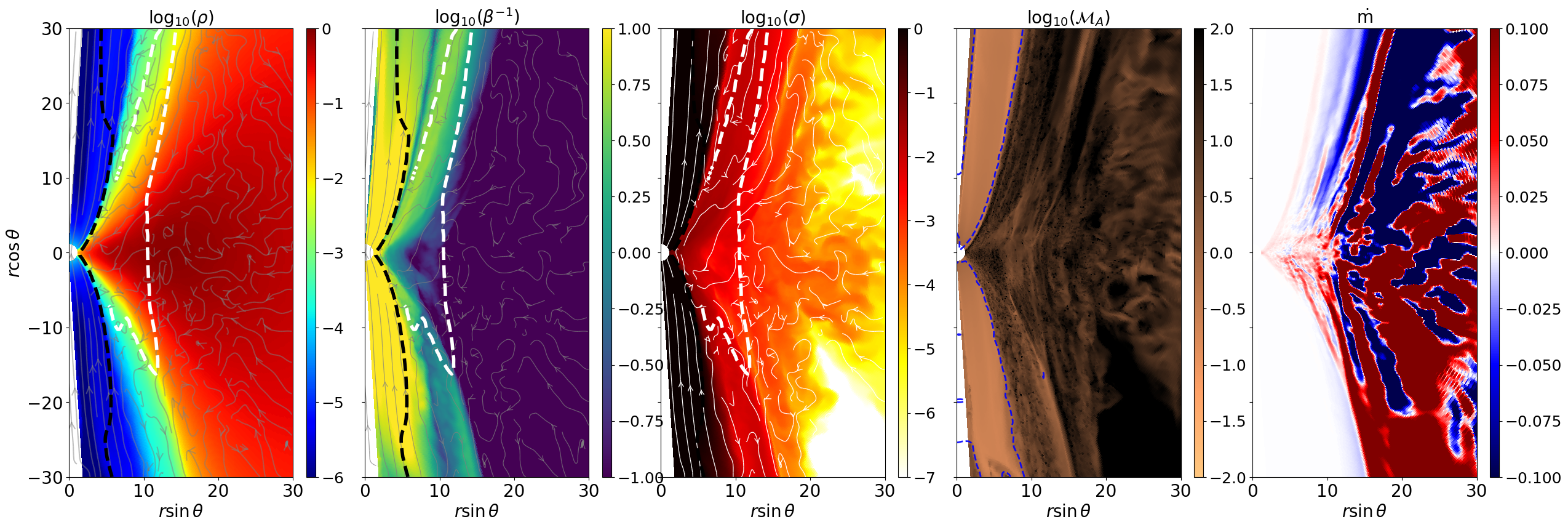} \\[1ex]

    \textbf{plasma $\beta_{0} = 0.007$} \\
    \includegraphics[width=\linewidth]{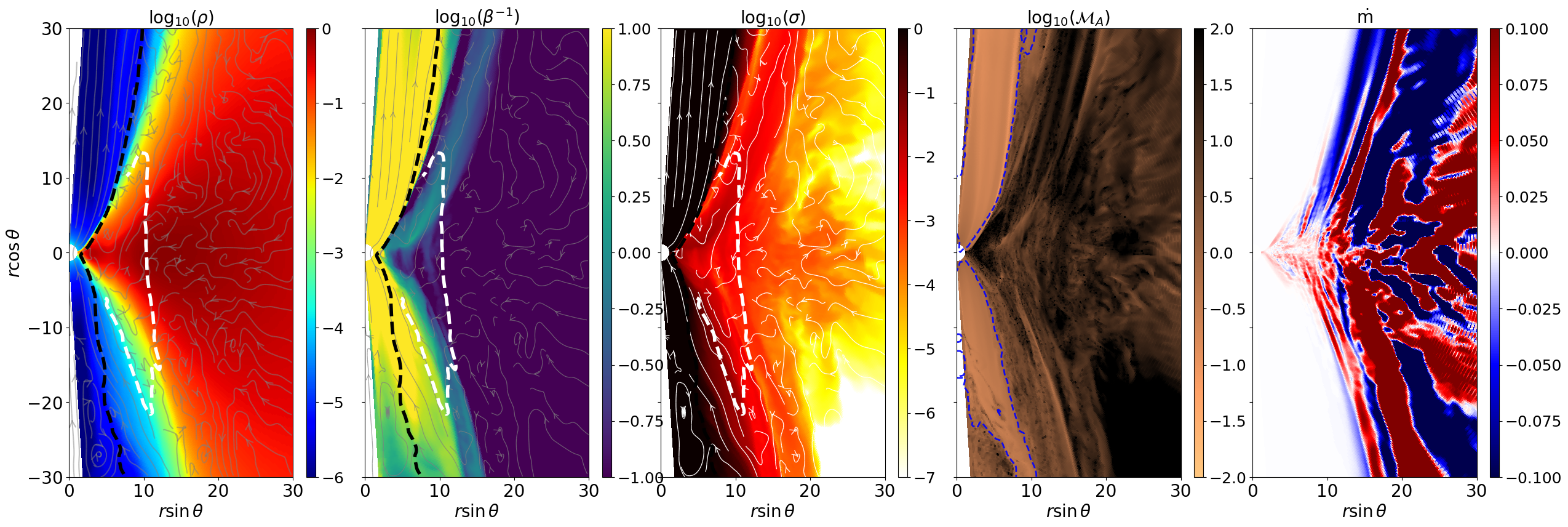}

    \caption{Same as Figure~\ref{fig:figure_4}, but for a fixed deformation of $30^\circ$, showing the variation with plasma-$\beta$. Each row corresponds to a different value of $\beta$: top—$\beta = 0.001$, middle—$\beta = 0.005$, and bottom—$\beta = 0.007$. All quantities are time-averaged over $t = 9000$–$11000\,t_{g}$ across the simulation domain (r–$\theta$ plane).}
    \label{fig:figure_5}
\end{figure*}

We present time-averaged 2D plots of various physical quantities to illustrate the evolution of the accretion column and the overall flow structure. The colormaps in Figures~\ref{fig:figure_4} and~\ref{fig:figure_5} capture key features such as disk-wind interaction, magnetization patterns, and accretion flow. We discuss these 2D maps in detail in the following sections: This section~\ref{sec-3.3} focuses on the variation with deformation, while Section~\ref{sec-3.4} explores the dependence on plasma $\beta$.

Figure~\ref{fig:figure_4} displays the time-averaged 2D distribution of density, $\beta^{-1}$, magnetization ($\sigma$), Alfv\'{e}n Mach number, and mass flux, $\dot{\rm{m}}$ for the fiducial case with $\beta = 0.005$, averaged over $t = 9000$–$11000$  $t_{g}$. Each row corresponds to a different deformation angle: $30^\circ$ (top), $45^\circ$ (middle), and $60^\circ$ (bottom). The black contour denotes the jet boundary, while the white contour indicates the wind boundary (see Section \ref{sec-2.3}). These boundaries are overlaid on the density, $\sigma$, and $\beta^{-1}$ plots. Magnetic field lines are shown as additional contours in these same plots.
The Alfv\'{e}n Mach number panel is computed as in e.g., \citet{2019ApJ...882....2V} 
\begin{equation}
    \mathcal M_{\rm A} = \frac{u_{p}}{c_{A}}, 
    \qquad 
    c_{A}^{2} = \frac{B_{p}^{2}}{\rho\,h + B_{p}^{2}} 
\end{equation}
where \(u_{p}\) is the poloidal four-velocity magnitude,  
\(u_{p}^{2} = u^{r}u_{r} + u^{\theta}u_{\theta}\),  
and \(B_{p}\) is the poloidal magnetic-field strength,  
\(B_{p}^{2} = B^{r}B_{r} + B^{\theta}B_{\theta}\),  
and \(h = \frac{\gamma}{\gamma-1}\,\frac{P}{\rho} + 1\) is the specific enthalpy used in our calculations.

The second column in Figure~\ref{fig:figure_4}, showing the $1/\beta$, highlights the pressure dominance within the flow. In this colormap, dark purple regions correspond to gas-pressure-dominated zones, while the green to yellow regions along the polar funnel indicate magnetically dominated areas. Collimated magnetic field lines are clearly visible along the polar direction for the $30^\circ$ and $45^\circ$ deformation cases in the density, $1/\beta$, and $\sigma$ plots, whereas the remaining field structure in the disk appears disordered and turbulent. 
The $\sigma$ in the third column not only traces the jet boundary but also reflects the magnetic field strength distribution, with higher values corresponding to stronger magnetization near the jet axis.

The fourth column in Figure~\ref{fig:figure_4} shows the Alfv\'{e}n Mach number in logarithmic scale, \(\log_{10} M_A\). Values with \(\log_{10} M_A < 0\) correspond to sub-Alfv\'{e}nic regions where magnetic forces dominate the flow and Alfv\'{e}n waves can propagate upstream along magnetic field lines, while \(\log_{10} M_A > 0\) indicates super-Alfv\'{e}nic regions where plasma motion exceeds magnetic disturbances, potentially forming shocks or discontinuities.  Sub-Alfv\'{e}nic regions are concentrated near the polar axis, but in the surrounding wind/jet sheath, \(\log_{10} M_A\) shows a mixture of sub- and slightly super-Alfv\'{e}nic values. The blue dashed contour in the figure marks \(M_A = 1\), highlighting the transition between magnetically dominated and flow-dominated regions.

The fifth column in Figure~\ref{fig:figure_4} presents the mass flux, $\dot{\rm{m}}$ = -$\sqrt{-g}\rho u^{r}$. In this plot, blue regions indicate negative values corresponding to outflows, while red regions represent positive values associated with inflows. In the mass flux maps, the polar region appears white because the density is very low there.

In Figure~\ref{fig:figure_4}, for the case with $\beta = 0.005$ and $30^\circ$ deformation, the magnetic field lines appear more collimated along the polar axis compared to the $45^\circ$ and $60^\circ$ cases. At $60^\circ$ deformation, both the $\beta^{-1}$ and $\sigma$ values are significantly reduced in the $+z$ direction, indicating a more gas-pressure-dominated flow in contrast to the more magnetically dominated structure at lower deformation angles. The strength and symmetry of the polar outflows also decrease with increasing deformation, following the trend: $30^\circ > 45^\circ > 60^\circ$. This is further supported by the mass flux maps, where the outflow component is strongest for the $30^\circ$ deformation, and progressively weaker at higher angles. 

\begin{figure}
    \centering
    \includegraphics[width=\linewidth]{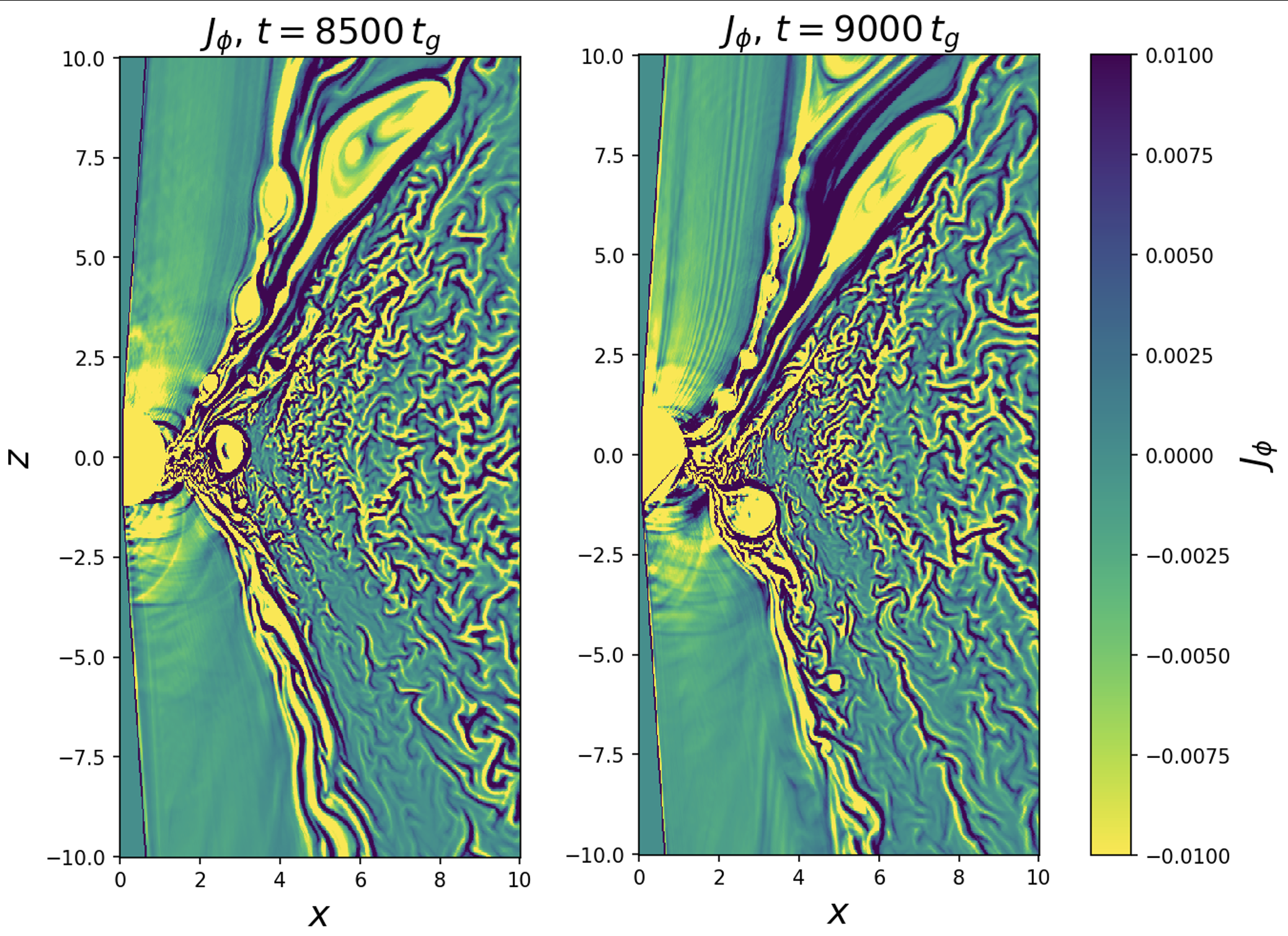}
    \includegraphics[width=\linewidth]{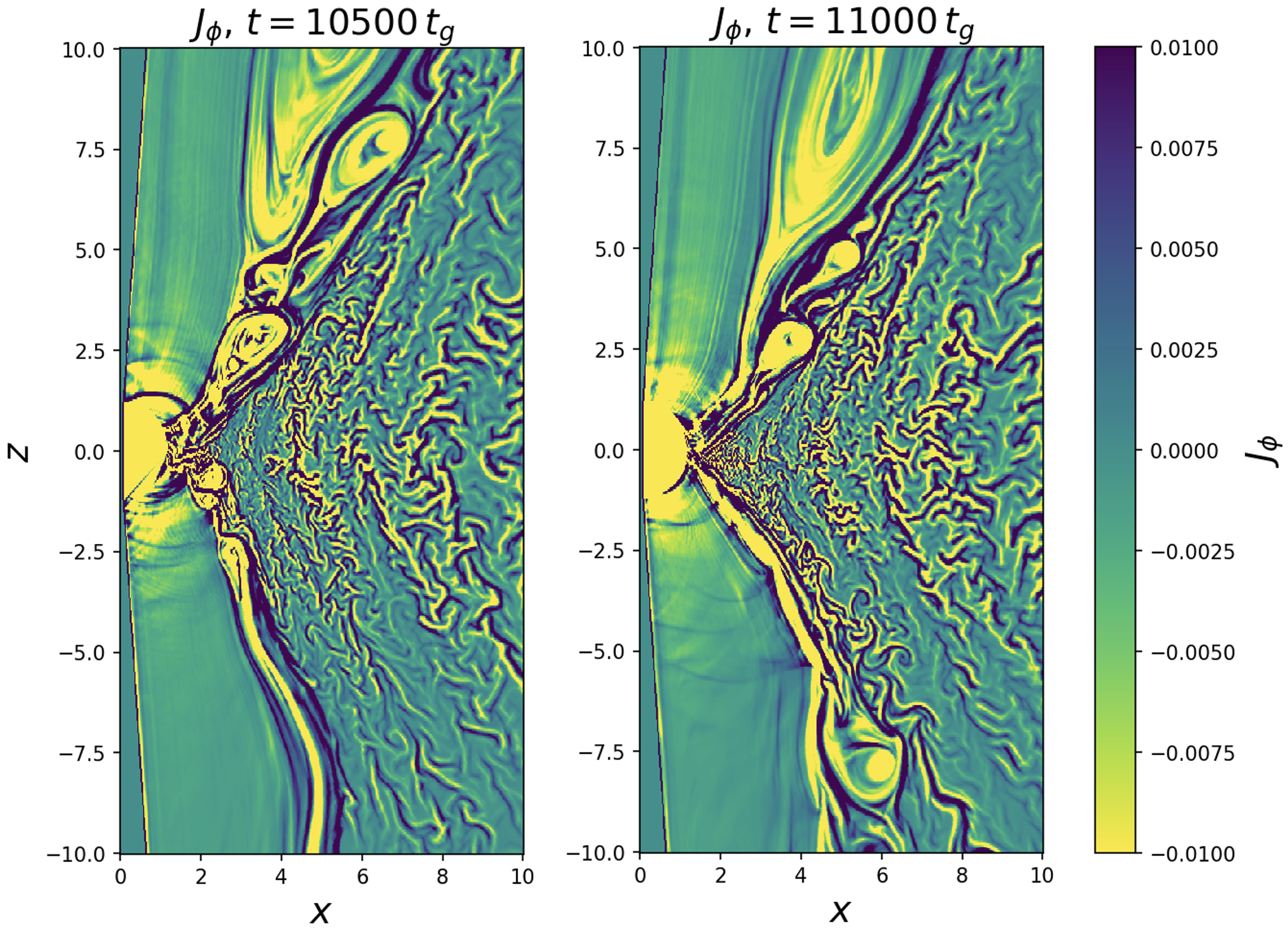}
    \caption{Azimuthal component of the electric current density, $J_\phi$, shown at four representative time snapshots ($t = 8500\,t_g$, $9000\,t_g$, $10500\,t_g$, and $11000\,t_g$) for the model with plasma $\beta = 0.001$ and an initial magnetic field deformation angle of $30^\circ$. The high-resolution simulation ($2048 \times 1024$) clearly captures the formation of thin current sheets, their subsequent fragmentation, and the development of magnetic reconnection sites accompanied by plasmoid formation.}
    %These features highlight the highly dynamic and non-linear evolution of the magnetized flow driven by strong magnetic stresses.
    \label{fig:figure_6}
\end{figure}

The variation in magnetic structure and outflow strength with deformation can be attributed to the underlying flow being in a SANE state, rather than a MAD configuration. In the absence of strong, coherent poloidal magnetic flux near the black hole, the system does not launch powerful, relativistic jets. Instead, we observe axisymmetric, magnetically-influenced winds emerging along the polar regions. 

At lower deformation angles (e.g., $30^\circ$), the magnetic field remains more ordered and collimated near the poles, facilitating stronger and more persistent polar winds. The higher values of $\beta^{-1}$ and $\sigma$ in these regions indicate moderate magnetization, sufficient to support outflows. As the deformation increases (e.g., to $60^\circ$), the magnetic field becomes more disordered, especially in the $+z$ direction, and the polar regions become increasingly gas-pressure dominated. This leads to weaker, less persistent outflows and lower mass loss rates, as also reflected in the mass flux maps. Overall, the trend, where outflow strength decreases as deformation increases $(30^\circ \rightarrow 45^\circ \rightarrow 60^\circ)$, is consistent with reduced magnetic coherence and wind efficiency at higher deformations.

%--------------------------------------------------------

\subsection{\MakeUppercase {Variation with plasma-$\beta$}}
\label{sec-3.4}

Figure~\ref{fig:figure_5} presents time-averaged 2D distributions of various quantities for different initial plasma-$\beta$ values: $\beta = 0.001$, $0.005$, and $0.007$, shown in the top, middle, and bottom rows, respectively, all at a fixed deformation angle of $30^\circ$. As \(\beta\) increases, corresponding to a weaker initial magnetic field, the jet sheath becomes less magnetized, and the magnetic field lines are less organized and less collimated.

For the lowest value, $\beta = 0.001$, the polar regions exhibit well-collimated magnetic field lines and a magnetically dominated funnel, conducive to structured, axisymmetric outflows. In contrast, at $\beta = 0.007$, the field becomes turbulent and disordered, especially along the polar axis, leading to a loss of collimation. This weak field regime results in outflows that are less magnetically regulated and more symmetric between inflow and outflow, as reflected in the mass flux map, showing a mix of inflow and outflow, unlike other cases.

Despite variations in magnetic strength, cases with $\beta = 0.001$ and $0.005$ show that the outflows are preferentially directed along the $+z$ direction, indicating a persistent large-scale asymmetry possibly seeded by initial conditions or disk-wind geometry. Stronger magnetic fields in the lower-$\beta$ cases help reinforce this asymmetry and guide the flow more effectively, while higher-$\beta$ cases show reduced magnetic control and more isotropic or turbulent behavior.

Figure~\ref{fig:figure_6} highlights reconnection-prone regions in our simulations through the spatial structure of the azimuthal component of current, $J_\phi \equiv (\nabla \times \mathbf{B})_\phi$. Snapshots from the high-resolution run reveal the formation of thin, elongated current sheets in the wind–sheath region, which subsequently fragment into localized, plasmoid-like structures. See Appendix~\ref{app_A} for details regarding the high-resolution run. The colormap shows the signed azimuthal current density, where positive and negative values indicate opposite toroidal current polarities. Regions exhibiting sharp polarity reversals correspond to strong current sheets. These current sheets extend asymmetrically into the funnel wall and wind outflow, breaking north–south symmetry. Such features are characteristic of reconnection in strongly magnetized, turbulent flows. The presence, location, and temporal evolution of these current sheets suggest that the reconnection-driven magnetic field plays an important role in asymmetric outflow dynamics.

%--------------------------------------------------------
\subsection{\MakeUppercase {Mass Accretion Rate and Relativistic outflows}}
\label{sec-3.5}
\begin{figure}
    \centering
    \includegraphics[width=\linewidth, height=8cm]{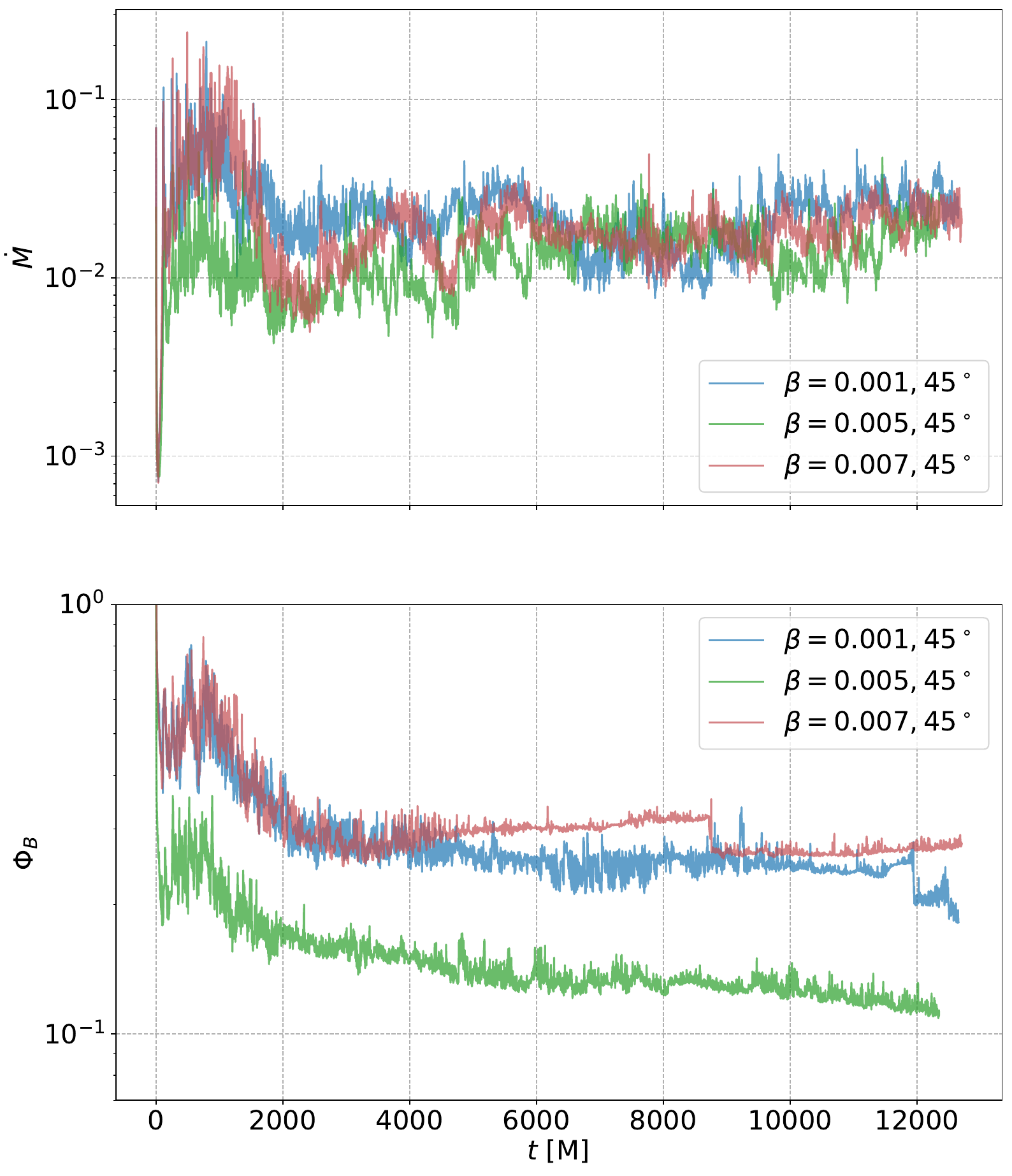}
    \caption{Time evolution of the mass accretion rate ($\dot{M}$, top panel) and dimensionless magnetic flux ($\Phi_{\rm BH}$, bottom panel) at the black hole horizon for three different initial magnetization levels: $\beta = 0.001$, $0.005$, and $0.007$ for 45 degrees.}
    \label{fig:figure_7}
\end{figure}

\begin{figure}
    \centering
    \includegraphics[width=\linewidth, height=8cm]{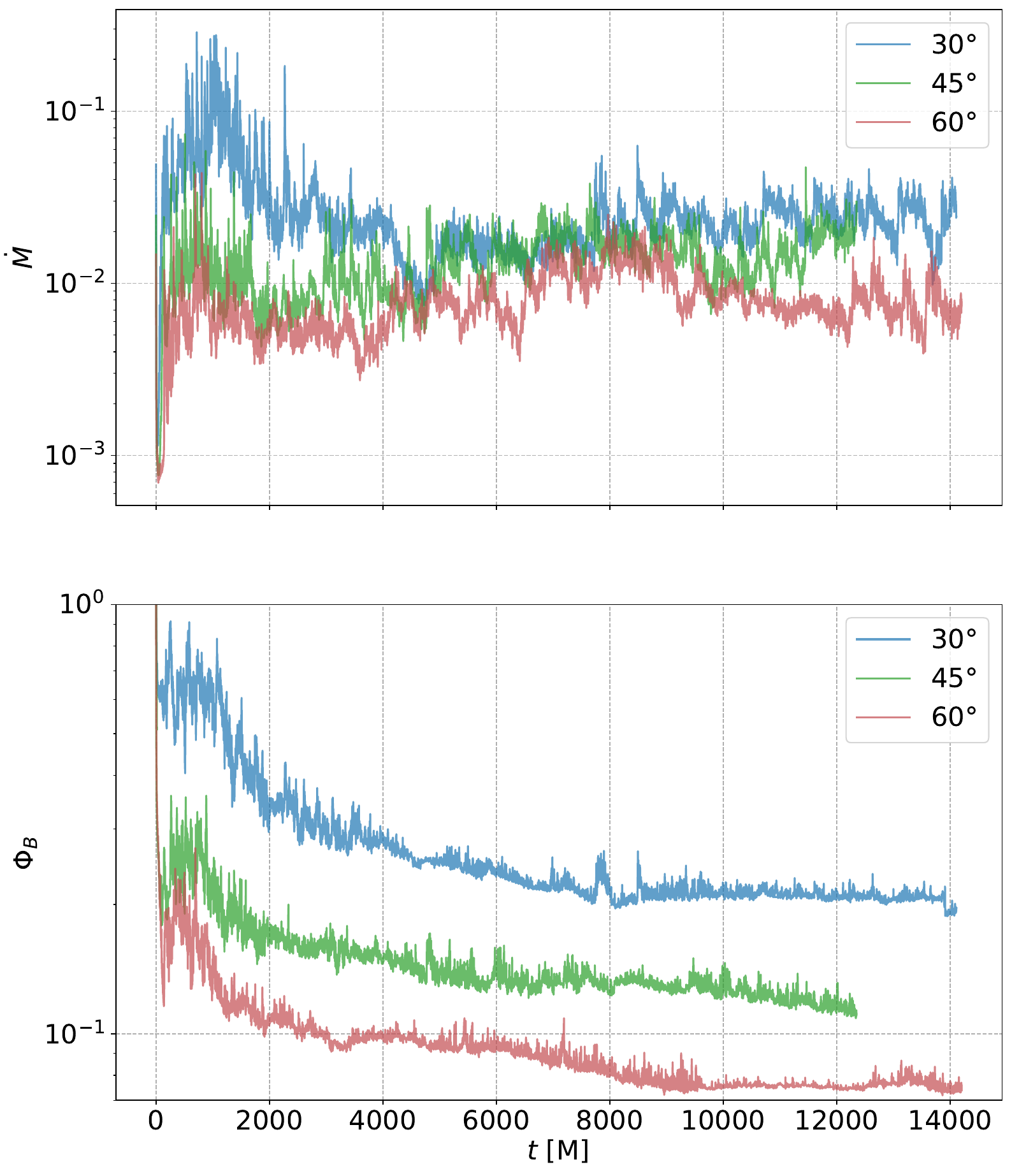}
    \caption{Same as Fig.~\ref{fig:figure_6} but for magnetic field deformation angles  of \( 30^\circ \), \( 45^\circ \), and \( 60^\circ \), at fixed \( \beta = 0.005 \).}
    \label{fig:figure_8}
\end{figure}

\begin{figure*}[ht]
    \centering
    \includegraphics[width=0.94\linewidth]{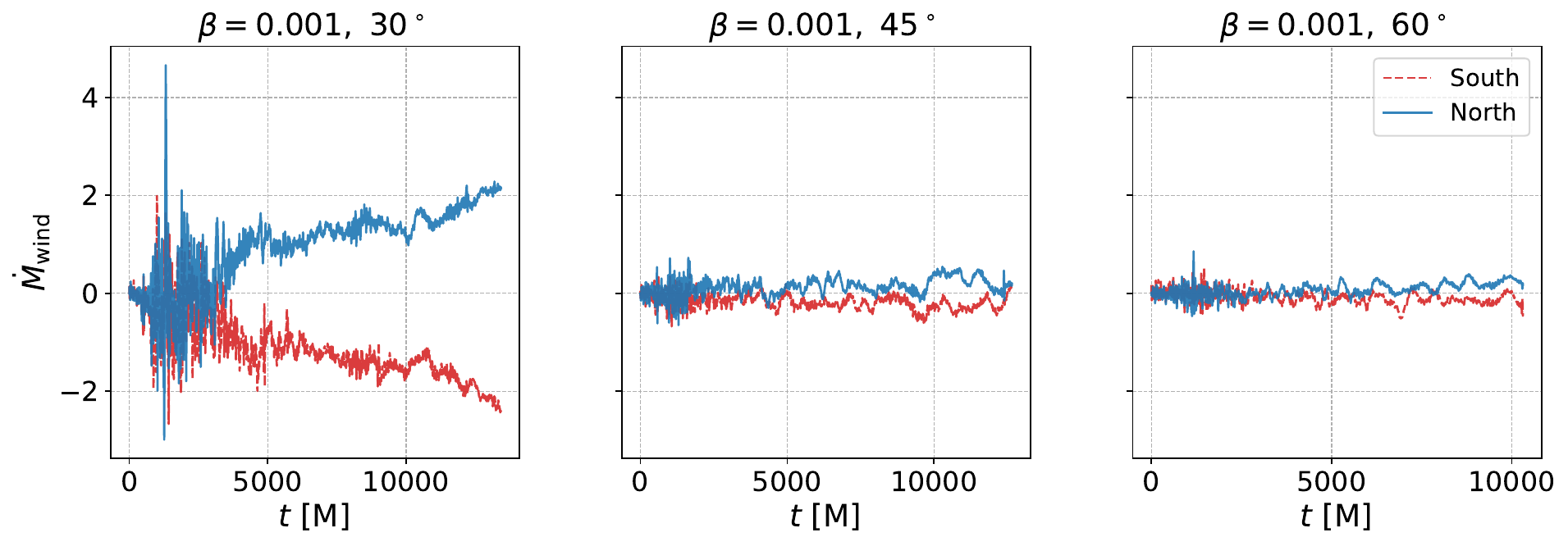}
    \includegraphics[width=0.94\linewidth]{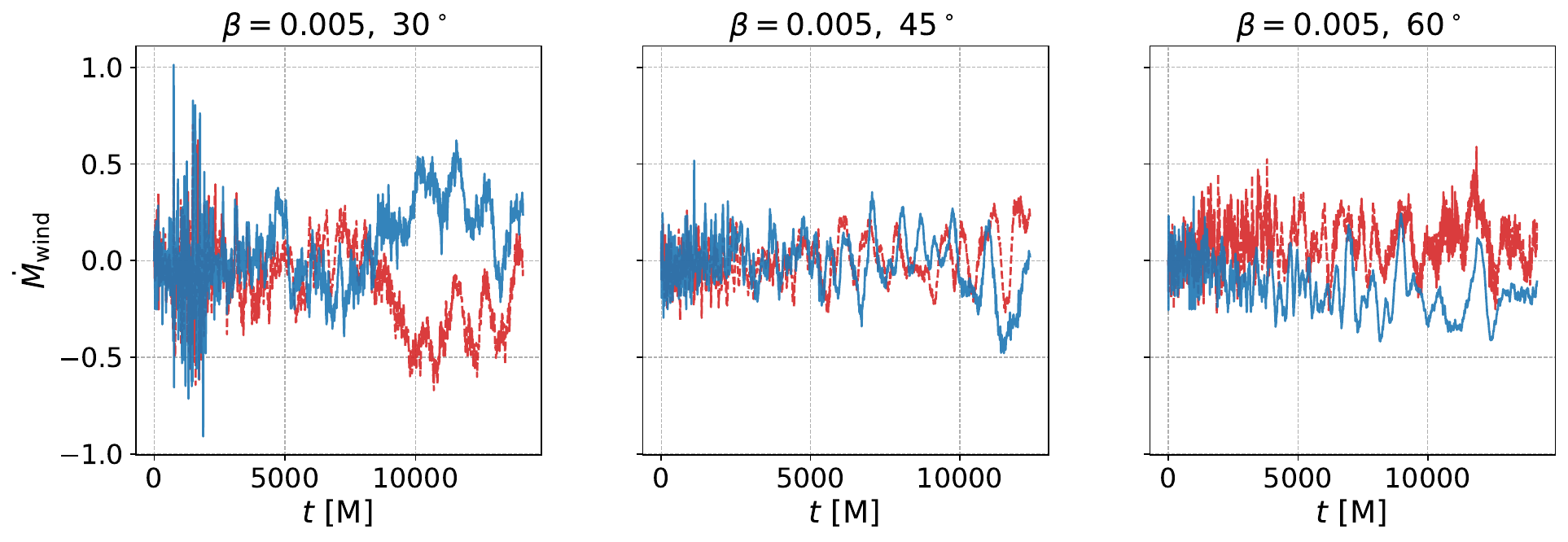}
    \includegraphics[width=0.94\linewidth]{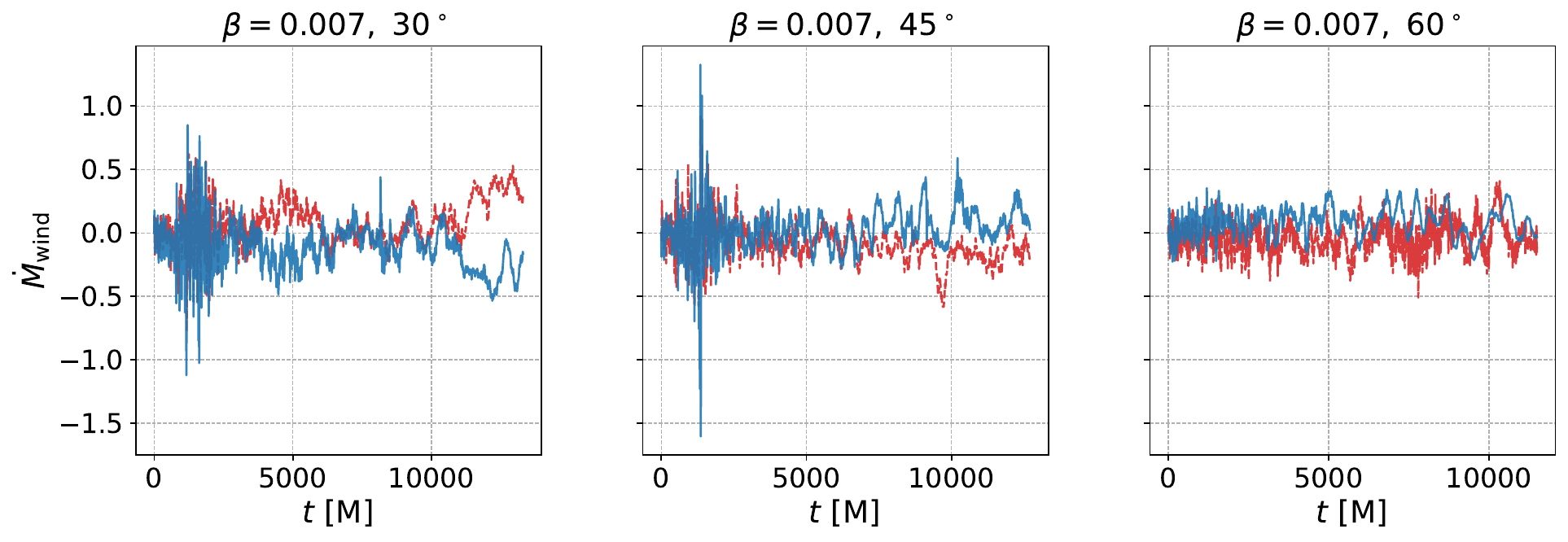}
    \caption{Time evolution of radial mass flux corresponding to the wind across different initial plasma-\(\beta\) values and deformation angles. Each panel shows results for a particular combination of \(\beta\) and deformation, with columns corresponding to deformations of \(30^\circ\), \(45^\circ\), and \(60^\circ\), and rows corresponding to \(\beta = 0.001\) (top), \(\beta = 0.005\) (middle), and \(\beta = 0.007\) (bottom). Blue and red lines represent the mass flux in the northern and southern hemispheres, respectively.}
    \label{fig:figure_9}
\end{figure*}

\begin{figure}
    \centering
    \includegraphics[width=\linewidth]{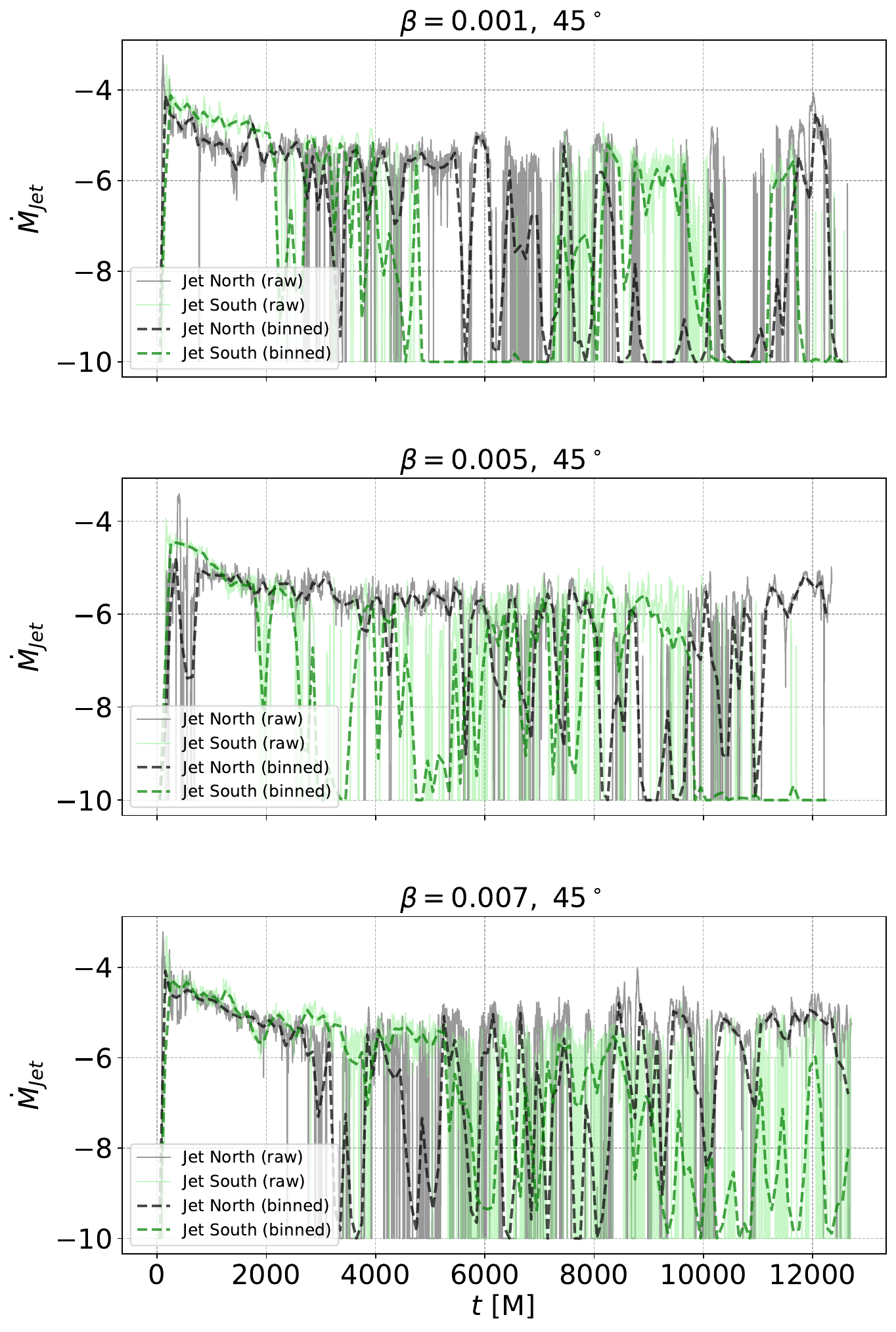}
    \caption{Comparison of jet power and properties between the northern and southern hemispheres as a function of time. The plot shows the evolution of mass fluxes of the jet component  in both directions for different initial magnetizations at an deformation of \( 45^\circ \).}
    \label{fig:figure_10}
\end{figure}

Understanding the time variability of the mass accretion rate onto the black hole is essential for characterizing the dynamics of the accretion flow and its coupling to jet production or wind outflows. In this section, we analyze the temporal behavior of \( \dot{M} \) and its relationship with magnetic flux \(\Phi_{\rm BH} \) across simulations with varying initial magnetization levels and deformation angles (magnetic flux normalization same as in \citet{2021MNRAS.506..741M}). We also examine the time evolution of the mass outflow rates, $\dot{M}_{\rm wind}$ and $\dot{M}_{\rm jet}$, calculated at $r = 50\,r_{\rm g}$, and analyze their behavior separately in the northern and southern hemispheres.

Figure~\ref{fig:figure_7} shows the time evolution of the mass accretion rate (\( \dot{M} \), top panel) and the dimensionless magnetic flux threading the black hole horizon (\( \Phi_{\rm BH} \), bottom panel). We compare the mass accretion rate for three different magnetizations: \( \beta = 0.001 \), \( 0.005 \), and \( 0.007 \), all at an deformation angle of \( 45^\circ \). The accretion rates appear comparable across all three values of $\beta$. However, the magnetic flux $\Phi_{\rm BH}$ is found to be highest for the weakest magnetization case ($\beta = 0.007$) and lowest for the intermediate case ($\beta = 0.005$).

This behavior highlights the non-trivial role of magnetic fields in regulating accretion. This ordering may be understood in terms of the interplay between magnetic field strength and wind/outflow dynamics. In the strongest field case ($\beta = 0.001$), early-time outflows may be vigorous enough to efficiently carry away magnetic flux, thereby limiting its long-term accumulation near the black hole. In contrast, for $\beta = 0.007$, the magnetic field is initially weaker, allowing a prolonged phase of flux accumulation before significant outflows develop. The comparatively weaker initial field in the $\beta = 0.007$ case may delay the onset of magnetically driven winds, enabling magnetic flux to build up to higher levels before any significant depletion mechanism begins.

The observed trends in \( \Phi_{\rm BH} \) for the $\beta = 0.007$ case show a sharp decrease around $t = 8500\,t_{g}$. This could result from saturation of the magneto-rotational instability (MRI), after which turbulent reconnection processes become dominant, leading to magnetic energy dissipation. As the system evolves, stronger outflows or winds may develop and begin to remove magnetic flux from the vicinity of the black hole. While in strongly magnetized flows (\(\beta = 0.007\)) the large magnetic tension can partially inhibit mass accretion, in more weakly magnetized flows (higher \(\beta\)) the magnetic support is weaker, allowing gas to accrete more efficiently. However, because the magnetic field is initially weaker, the net magnetic flux accumulated near the black hole remains lower compared to the strongly magnetized case. Thus, the observed trends in \( \dot{M} \) and \( \Phi_{\rm BH} \) for (\(\beta = 0.007\)) case. reflect a competition between gravitational infall and magnetic pressure, with important implications for relativistic outflows and energy extraction in accreting systems.

Figure~\ref{fig:figure_8} presents the time evolution of the mass accretion rate \( \dot{M} \) for three different magnetic field's deformation angles, \( 30^\circ \), \( 45^\circ \), and \( 60^\circ \), at a fixed plasma beta \( \beta = 0.005 \). The \( 30^\circ \) deformation case shows a consistently higher accretion rate compared to the more inclined geometries. This suggests that a stronger vertical component of the magnetic field in lower deformation cases allows more matter to be funneled efficiently toward the black hole. This can be seen in 2D colormaps of mass flux in Figure~\ref{fig:figure_4}.

Interestingly, the \( 30^\circ \) case also maintains relatively strong magnetic flux levels, indicating that the field lines remain well-coupled to the inflowing gas, thereby facilitating efficient accretion. As the deformation increases, the magnetic field geometry becomes more distorted. This is particularly pronounced for higher deformation angles, following the trend \( 60^\circ > 45^\circ > 30^\circ \), and is indicative of turbulent reconnection and disruption in the outflow-launching region. As a result, the structure and strength of the northward jet are compromised, leading to a decrease in mass accretion rate with increasing deformation.

Figure~\ref{fig:figure_9} presents the time evolution of radial mass flux corresponding to the wind with blue and red representing the flows in the northern and southern hemisphere, respectively. When the blue/red line is above zero, it indicates outflowing mass in the corresponding hemisphere; when it dips below zero, it indicates temporary inflow or fallback.

Simulations with lower plasma \(\beta\) exhibit stronger outflows in the northern hemisphere, whereas the southern hemisphere often shows sustained inflow. This hemispheric asymmetry becomes more prominent with decreasing \(\beta\) and deformation angle, indicating that stronger magnetic fields (i.e., lower \(\beta\)) may promote more efficient wind launching along preferred directions. The magnetic pressure gradient can channel the outflows anisotropically, especially when magnetic field lines are more ordered and anchored closer to the pole.

This asymmetry is most pronounced in the case with \(\beta = 0.001\) and an deformation angle of \(30^\circ\), where strong, magnetically-driven outflows dominate the northern hemisphere while the southern side remains inflow-dominated. Conversely, for \(\beta = 0.007\) and an deformation angle of \(60^\circ\), the outflow is nearly symmetric and weak, suggesting that weaker magnetic fields and more equatorially asymmetric field configurations are less effective at producing directional winds. The general variability and imbalance between hemispheres highlight the role of magnetic pressure and field geometry in shaping the global wind structure.  Magnetic pressure can provide additional support against gravity and drive asymmetric mass ejection, particularly in regions where the field is stronger or more ordered. Additionally, the geometry of the field lines, whether more aligned with the rotation axis or more tangled due to turbulence, affects how and where outflows are launched. These factors together contribute to the observed north-south asymmetries in the wind, especially in simulations with stronger magnetic fields (i.e., lower \(\beta\)) and lower deformation angles.

Figure~\ref{fig:figure_10} shows the time evolution of radial mass flux of jet component in the northern (light gray) and southern (light green) hemispheres for simulations with different initial magnetization at \(45^\circ\) deformation. The jet activity appears intermittent, with episodic enhancements in both directions, suggesting the presence of very weak/ transient jet launching events. Notably, the northern jet tends to be more prominent across most cases, although both hemispheres exhibit variability likely influenced by the initial plasma \(\beta\) and magnetic reconnection processes near the black hole. 

Lower plasma $\beta$ values imply stronger magnetic fields, which can support sustained jet launching but also amplify asymmetric reconnection-driven fluctuations. Reconnection events in the turbulent disk and funnel wall can temporarily disrupt the magnetic structure needed for steady jet collimation, leading to the observed bursty, hemispherically asymmetric outflows. 

To further investigate the weak/episodic nature of the jet activity observed in Figure~\ref{fig:figure_10}, we generate time-averaged jet mass flux for visual comparison. In Figure~\ref{fig:figure_10}, the data binned over fixed time intervals of width 100 $t_{g}$ are shown in a bold gray line for the northern counterpart and a bold green line for the southern counterpart. This binning smooths short-term fluctuations and reveals the underlying long-term behavior of the jet power. The raw curves, shown in light gray (north) and light green (south), display fluctuations, indicating that the jet activity is not steady but rather episodic or intermittent. This broken or bursty appearance suggests that the jet launching process may be modulated by time-dependent physical processes such as magnetic instabilities, turbulence, or variable accretion. In particular, the northern jet typically exhibits a more persistent, coherent outflow, while the southern counterpart appears more irregular and weaker, especially at lower magnetization levels. The binning process thus provides a clearer picture of the jet dynamics, helping to distinguish genuine episodic behavior from high-frequency variability.

The relatively weak or absent jets in our simulations are consistent with a SANE accretion state, where the magnetic flux near the black hole is not sufficient to launch strong, persistent jets. Similar to findings by \citet{2025PhRvD.112f3013C}, even in misaligned or weakly magnetized flows, jet launching is limited, and local flux redistribution caused by equatorial magnetic field deformations can further suppress sustained jet activity.

We note that the strength of the BZ jet can, in principle, be sensitive to the choice of density floor and magnetization limits, which are required for numerical stability in GRMHD simulations. In our implementation, these floor values are maintained as close as possible to standard prescriptions, ensuring that sufficiently high magnetization ($\sigma \gtrsim 1$) is preserved in the polar regions. We find that the weak and intermittent jet activity observed here cannot be attributed solely to the floor treatment. Instead, the imposed magnetic-field deformation leads to asymmetric wind-driven outflows that redistribute magnetic flux away from the black hole spin axis into equatorial, mass-loaded regions, where the field tends to close or reconnect, reducing the amount of open horizon-threading flux available for steady BZ jet launching. This indicates that the diminished BZ power is a non-trivial consequence of the asymmetric magnetic geometry and outflow dynamics rather than a numerical artifact. See Appendix~\ref{app_B} for MRI quality details.

%--------------------------------------------------------
\subsection{\MakeUppercase {Power Spectral Density and Power-Law Fitting}}
\label{sec-3.6}
To study the variability characteristics of the accretion flow, we compute the Power Spectral Density (PSD)- the distribution of signal power as a function of frequency of the mass accretion rate \( \dot{M} \) using a Fourier transform of the time series data. This analysis is carried out for the case with plasma beta \( \beta = 0.001 \) and a magnetic field deformation angle of \( 30^\circ \). Among all the simulated runs, this configuration displays the highest degree of asymmetry in the outflow (see Figure~\ref{fig:figure_9}). All PSD calculations are carried out assuming a black hole mass of $10 M_\odot$.

The resulting PSD exhibits a broadband noise profile, indicative of stochastic turbulence in the accretion flow. No dominant, coherent oscillatory modes are detected over the full simulation duration, suggesting the absence of long-lived periodicities. The PSD shown here corresponds to this time interval, $t = 4{,}000{-}12{,}000\,t_{g}$.

\begin{figure}[ht]
    \centering
    \includegraphics[width=\linewidth, height=8cm]{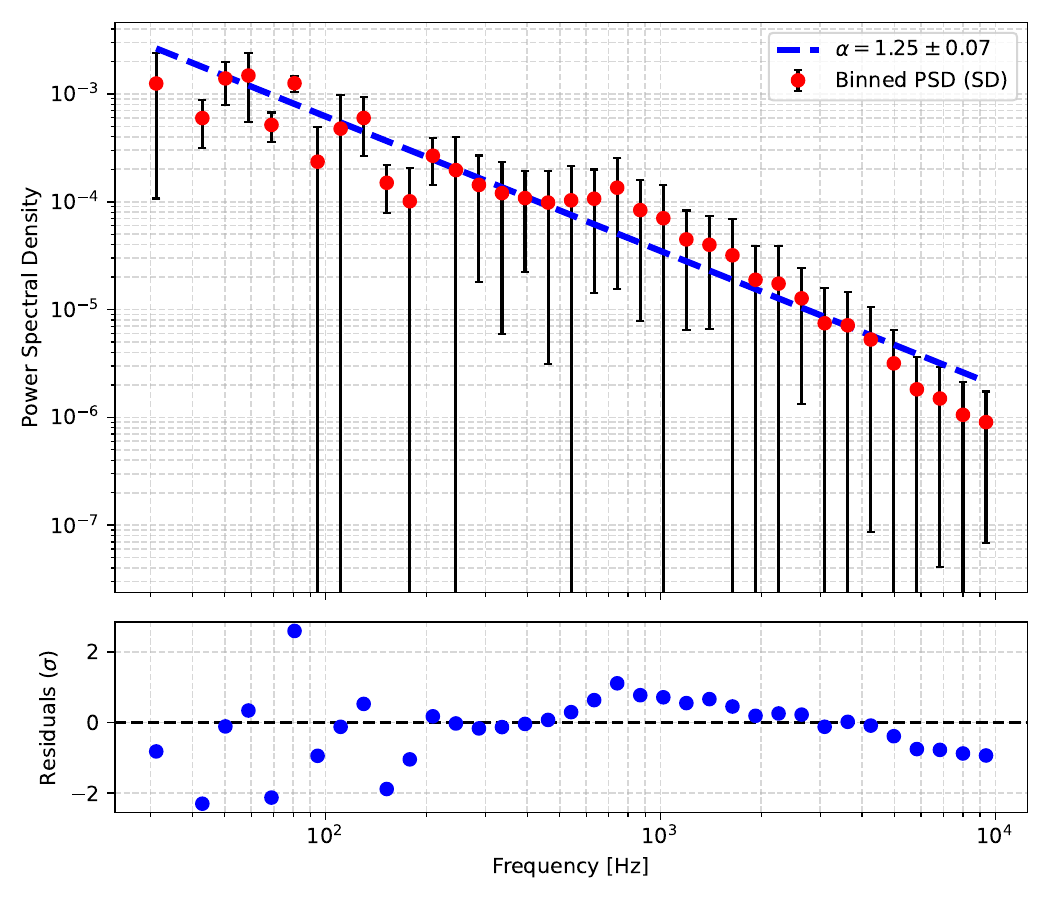}
    \caption{Power-law fitting of the binned PSD. The \textbf{top panel} shows the best-fit power-law model overplotted on the binned PSD data. The \textbf{bottom panel} displays the residuals of the fit, highlighting deviations from ideal power-law behavior and assessing the goodness of fit.}
    \label{fig:figure_11}
\end{figure}

To further reduce the high-frequency noise and reveal the underlying structure of the PSD, we apply logarithmic binning in frequency space. This approach smooths out rapid fluctuations while preserving the overall spectral shape, thereby enabling clearer identification of characteristic trends, power-law slopes, and any possible spectral breaks. The binned PSD is particularly useful for subsequent fitting procedures and interpretation of accretion flow dynamics across timescales.

Figure~\ref{fig:figure_11} shows the power-law fitting applied to the binned PSD of mass accretion rate \( \dot{M} \). The top panel overlays the best-fit power-law model on the binned PSD, while the bottom panel presents the residuals from this fit. The mass accretion rate variability follows a power law, \( P(f) \propto f^{-k} \), with \( k \) typically between 1 and 2 for all our models (see Table~\ref{tab:psd_k}). This range is consistent with magnetorotational instability (MRI)-driven turbulence, while a deficit of high-frequency power would suggest weaker MRI activity or a more laminar flow regime \citep{2025MNRAS.536.3284U}. The resulting power-law slope  for the most asymmetric model (\( \beta = 0.001 \), \( i = 30^\circ \)) presented in Figure~\ref{fig:figure_10} is approximately \( -1.25 \), which is indicative of MRI-driven turbulence.

This slope is broadly consistent with theoretical expectations for MRI-driven accretion flows, where turbulence generated by the MRI leads to stochastic fluctuations across a range of timescales. The extracted slope can be used to infer the level of turbulent activity and is often related to the effective viscous parameter \( \alpha \). Studies like \citet{2010MNRAS.408..752P}, suggest that steeper PSD slopes (i.e., more negative) generally correspond to more vigorous turbulence and higher effective \( \alpha \)-values. A slope of \( \sim -1.2 \) is indicative of moderate to strong MRI turbulence, consistent with the high magnetization (\( \beta = 0.001 \)) and the development of sustained accretion variability in this simulation setup.

The power-law indices corresponding to all other models, fitted to their PSDs over the time range $t = 4000$ to $12000\,t_g$, are provided in Appendix~\ref{app_C} (see Table~\ref{tab:psd_k}). The power-law indices lie in the range \( k \approx 1.17{-}1.42 \), showing only modest variation with deformation but a slight steepening trend with decreasing plasma-\( \beta \).

%-------------------------------------------------------------

\section{Discussion}
\label{sec-4}

\subsection{\MakeUppercase {Connecting to Observations}}

Our simulations demonstrate that asymmetric winds naturally emerge from a combination of magnetic field geometry and its deformation. The corresponding PSD analysis offers insight into the variability mechanisms: steep slopes (\(k>2\)) indicate large-scale, quasi-steady magnetic structures, while flatter slopes (\(k \sim 1\)) reflect stochastic, turbulent variability driven by small-scale accretion and wind-launching dynamics \citep{2005MNRAS.359..345U}.  

The high-frequency slope of the simulated mass accretion rate ($\dot{M}$) in Figure~\ref{fig:figure_11}, $k \simeq -1.25$, is indicative of red-noise variability arising from magnetically driven, asymmetric winds and episodic outflows near the inner accretion flow. However, direct quantitative comparison with observed X-ray PSDs of AGN or X-ray binaries is limited. Our simulations are axisymmetric (2D) and cannot capture true 3D oblique effects such as precession or non-axisymmetric instabilities. As a result, the simulated asymmetries primarily reflect equatorial and hemispheric variations rather than a fully 3D magnetosphere. Furthermore, the flow asymmetries are inferred indirectly from morphology and energetics, making direct observational diagnostics (e.g., polarization asymmetries or hemispheric jet power differences) challenging and model-dependent.  

Despite these limitations, the qualitative trends provide useful physical insight. The emergence of asymmetric winds, the variability patterns in $\dot{M}$, and the magnetic flux evolution near the horizon suggest that the equatorial asymmetry or deformation in magnetic fields can strongly influence relativistic outflows. Observationally, similar asymmetries are seen in NGC~7469, which exhibits a one-sided, stratified outflow up to $\sim$1700 km s$^{-1}$ \citep{2023ApJ...942L..37A}, while rapid jet reorientation is observed in microquasars such as GRS~1915+105 \citep{2025ApJ...986..108R}, and structured winds appear in AGN like Mrk~509 \citep{2013IAUS..290...45K}. These examples suggest that coupling between the inner accretion flow, magnetic field geometry, and outflow dynamics can naturally produce asymmetric winds and variability.

Our appendices further support the effects of deformation in magnetic field geometry. Appendix~ \ref{app_D} shows that the horizon flux retains a subtle imprint of the initial magnetic field deformation, never fully relaxing to a symmetric split-monopole. This indicates that even in 2D, the black hole partially preserves memory of the initial field geometry. Appendix~\ref{app_A} demonstrates that low- and high-resolution runs for the same model ($\beta = 0.001$, $30^\circ$ deformation) produce very similar qualitative behavior in winds, accretion rate, and magnetic flux, confirming that the main features of asymmetric outflows and variability are robust against resolution. Future 3D simulations are needed to more accurately capture obliquity effects, turbulence, and jet physics, and to enable more direct comparison with observations.

%----------------------------------------------------
\section{Conclusions}
\label{sec-5}

We have presented a comprehensive axisymmetric GRMHD study exploring how magnetic field strength (plasma-$\beta$) and deformation affect the structure, dynamics, and variability of black hole accretion flows and associated outflows. Through time-averaged 1D and 2D diagnostics, we find that both increased magnetization (lower-$\beta$) and higher deformation angles enhance relativistic outflows, jet collimation, and magnetic-pressure dominance, particularly in the northern hemisphere. Strong asymmetries persist across all simulations, with more efficient accretion and directional outflows occurring at low deformation ($30^\circ$) and strong field strength ($\beta = 0.001$).

Jet/Wind launching is influenced by both the magnetic field configuration and plasma conditions. In cases with relatively stronger magnetization, turbulence and reconnection activity appear reduced in the southern hemisphere, accompanied by more organized, collimated jet-sheath wind in the north. The simulations with stronger magnetic fields exhibit enhanced turbulence, increased reconnection, and greater asymmetries between hemispheres. The time evolution of radial mass flux corresponding to wind at large radii reflects this imbalance, with more persistent asymmetric outflow and temporal variability in cases where magnetic forces play a stronger role in shaping the outflows, though these remain far from a magnetically arrested state. deformation-dependent changes in field geometry also impact wind strength and directionality; at higher deformation angles, the distortion of magnetic field lines reduces the coherence of the outflows, resulting in more symmetric, less collimated structures.

Analysis of the mass accretion rate variability using PSD reveals red-noise behavior consistent with MRI-driven turbulence. The most magnetized, low-deformation case exhibits the steepest PSD slope ($\sim -1.25$), indicating strong turbulent activity and a dynamic, variable flow environment.

Altogether, these findings highlight the intricate interplay between magnetic field strength, geometry, and plasma dynamics in shaping black hole accretion flows, with significant implications for jet formation, wind launching, and variability in astrophysical systems.

\section*{Acknowledgment}
The simulations were performed using computational resources provided by the Center for Informatics and Computation in Astronomy (CICA) at National Tsing Hua University, funded by the Ministry of Education (MOE) of Taiwan. I.P. acknowledges support from the National Science and Technology Council (NSTC) of Taiwan under Grant No. 112-2811-M-007-055-MY3. I.K.D. acknowledges financial support from the TDLI Postdoctoral Fellowship. Y.M. acknowledges support from the National Key R\&D Program of China (Grant No.\,2023YFE0101200), the National Natural Science Foundation of China (Grant Nos.\,12273022, 12511540053), and the Shanghai Municipality Orientation Program of Basic Research for International Scientists (Grant No.\,22JC1410600). H.Y.K.Y. acknowledges support from the NSTC of Taiwan (NSTC 112-2628-M-007-003-MY3; NSTC 114-2112-M-007-032-MY3) and the Yushan Scholar Program of the MOE of Taiwan (MOE-108-YSFMS-0002-003-P1 \& MOE-114-YSFMS-0002-002-P2 ).

%\section{Comments}
\appendix
%\section{Appendix }
%\subsection*{\MakeUppercase {A. MRI Quality factor- Jet Dynamics}}
\section{Resolution Comparison}
\label{app_A}
\begin{figure}[htbp]
\centering
\begin{minipage}[b]{0.5\linewidth}
    \centering
    \includegraphics[width=\linewidth]{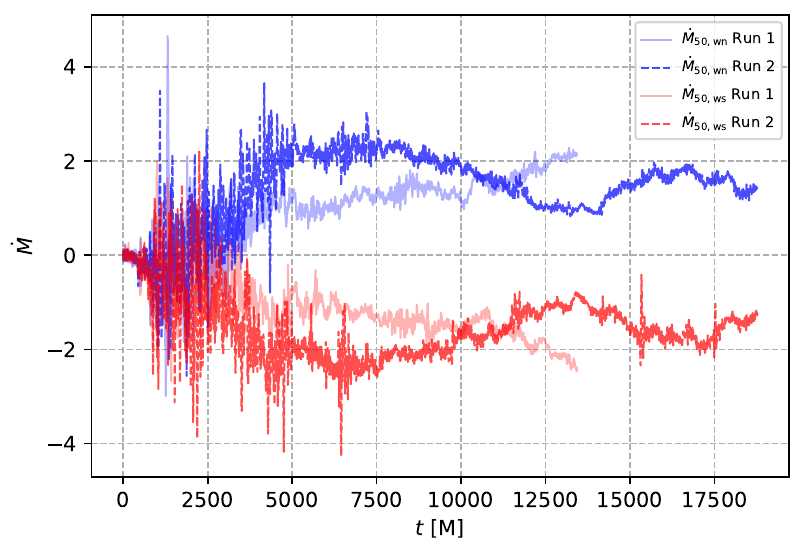}
    \caption{Comparison of northern and southern wind mass fluxes ($\dot{M}_{50,\mathrm{wn}}$ and $\dot{M}_{50,\mathrm{ws}}$) for the low-resolution (Run 1, faded, solid lines) and high-resolution (Run 2, bold, dashed lines) simulations.}
    \label{fig:wind_comparison}
\end{minipage}
\hfill
\begin{minipage}[b]{0.48\linewidth}
    \centering
    \includegraphics[width=\linewidth]{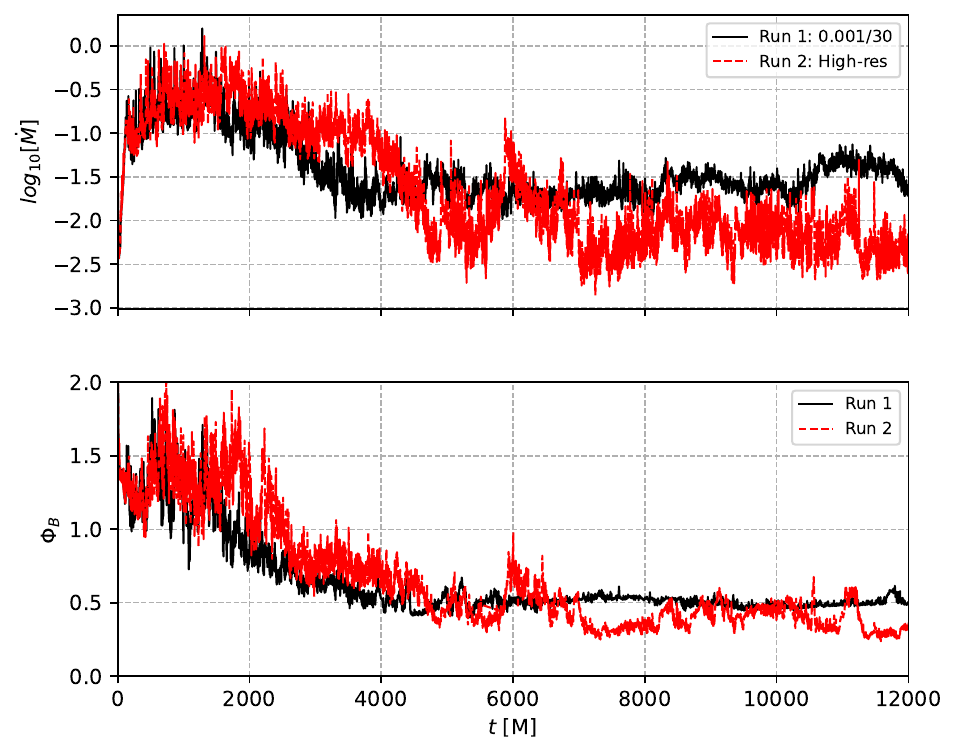}
    \caption{Time evolution of total mass accretion rate ($\dot{M}$, top) and magnetic flux $\Phi_B$ near the horizon (bottom) for the two simulations. Solid lines: Run 1 (low resolution), dashed lines: Run 2 (high resolution).}
    \label{fig:mdot_phi_compare}
\end{minipage}
\end{figure}

We compare the results of two simulations with the same initial magnetic field configuration ($\beta = 0.001$, deformation $30^\circ$) but different resolutions: a lower-resolution run ($1024 \times 512$) and a higher-resolution run ($2048 \times 1024$). Figure~\ref{fig:wind_comparison} shows the mass fluxes in the northern and southern winds ($\dot{M}_{50,\mathrm{wn}}$ and $\dot{M}_{50,\mathrm{ws}}$), while Figure~\ref{fig:mdot_phi_compare} presents the total mass accretion rate and magnetic flux near the horizon for these two runs. 

Overall, the two runs display very similar behavior, with comparable variability patterns, amplitudes, and trends in both the wind and accretion flow properties. This indicates that the main qualitative features of asymmetric winds, accretion rate variability, and magnetic flux evolution are robust against changes in resolution for this model. The higher-resolution run shows slightly enhanced fine-scale variability, but no major differences in the overall dynamics are observed.

\section{MRI Quality factor- Jet Dynamics}
\label{app_B}

The MRI, which is a crucial mechanism for angular momentum transport in accretion disks, drives turbulence in the magnetized flow, amplifying magnetic fields and facilitating the formation of strong outflows. This instability is responsible for the non-axisymmetric features observed in the radial and polar fluxes \citep{1991ApJ...376..214B}.

MRI enables outward angular momentum transport in differentially rotating accretion disks permeated by weak magnetic fields. This linear instability amplifies seed magnetic fields and disrupts laminar flow, triggering turbulence. The resulting turbulent shear stress serves as the dominant mechanism for radial angular momentum transport in our SANE models.

To ensure that the fastest-growing MRI mode is adequately resolved, we calculate the MRI quality factors $Q_{r}$ and $Q_{\theta}$, following the criteria discussed in \citet{2012MNRAS.426.3241N}, \citet{2019ApJS..244...10R}, etc. 
These quality factors represent how well the simulation grid resolves the characteristic wavelength of the MRI. 
In previous simulations for SANE accretion states, it has been shown that $Q_{\theta} \sim 5$--$10$, indicating sufficient resolution of the instability \citep{2004ApJ...605..321S}.

We have performed a higher-resolution simulation with a grid of $2048 \times 1024$, compared to the previous $1024 \times 512$. This enhanced resolution leads to significantly higher $Q_{\theta}$ values across the disk, including at $r>3$, demonstrating that the MRI is well resolved throughout the domain and that turbulent magnetic stresses efficiently transport angular momentum.

\begin{figure}
    \centering
    \includegraphics[width=0.6\linewidth]{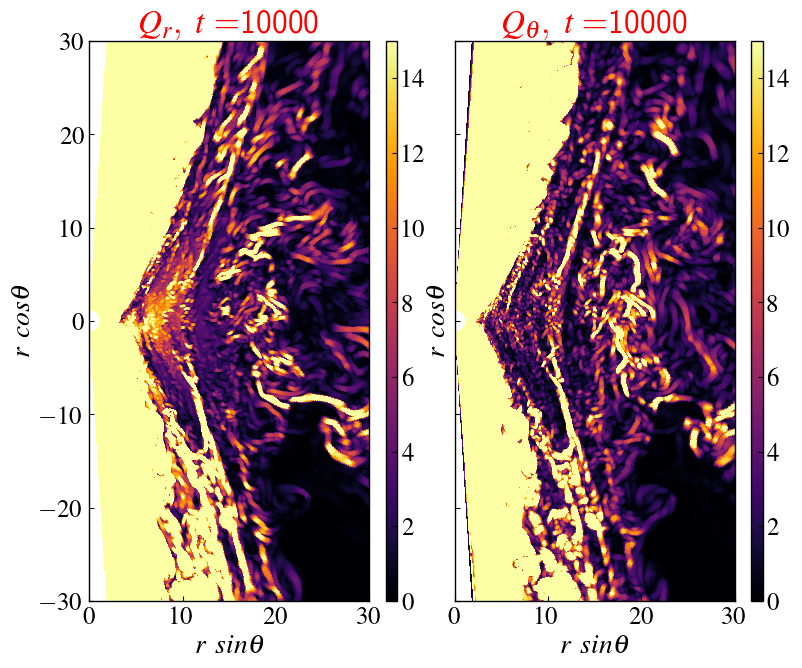}
    \caption{2D distribution of MRI quality factors for the simulation with \( \beta = 0.001 \) and a magnetic field deformation of \( 30^\circ \) at $t = 10,000\,t_{g}$. The resolution for this run is 2048 $\times$ 1024. The left panel shows \( Q_r \), and the right panel displays \( Q_\theta \). The higher values indicate a well-resolved MRI activity region.}
    \label{fig:figure_14}
\end{figure}

Figure~\ref{fig:figure_14} shows the 2D distribution of MRI quality factors for the simulation with \( \beta = 0.001 \) and a magnetic field deformation of \( 30^\circ \) at $t = 10,000\,t_{g}$. Elevated values of the MRI quality factors are observed, indicating that the fastest-growing MRI modes are well resolved there. 

Moreover, the enhanced MRI activity in the polar and near-jet regions leads to magnetic field collimation, which can contribute to the launching of magnetically driven winds and outflows in SANE accretion states.

\begin{table}[ht]
\centering
\caption{Power-law slopes $\alpha$ of the spectral density (PSD) computed over the time range $t = 8000\,t_g$ to $12000\,t_g$, for different values of plasma $\beta$ and deformation angles. The high-resolution run (30°, $\beta = 0.001$) is included at the end.}

\begin{tabular}{c|c|c|c}
\hline
deformation & \boldmath{$\beta = 0.001$} & \boldmath{$\beta = 0.005$} & \boldmath{$\beta = 0.007$} \\
\hline
30$^\circ$ & $1.25 \pm 0.07$ & $1.59 \pm 0.06$ & $1.59 \pm 0.06$ \\
45$^\circ$ & $1.28 \pm 0.06$ & $1.20 \pm 0.05$ & $1.46 \pm 0.07$ \\
60$^\circ$ & $0.76 \pm 0.06$ & $1.36 \pm 0.06$ & $0.87 \pm 0.06$ \\
30$^\circ$ (high-res) & $0.92 \pm 0.06$ & --- & --- \\
\hline
\end{tabular}
\label{tab:psd_k}
\end{table}

%\section*{Appendix B}
%\subsection*{\MakeUppercase{B. PSD for all models}}
\section{PSD for all models} 
\label{app_C}

Table~\ref{tab:psd_k} presents the variation of the power-law index, $k$, of the PSD, computed over the time interval $t = 8000,t_g$ to $12000,t_g$, for different values of plasma $\beta$ and deformation angles. The PSDs were logarithmically binned, and the slope uncertainties were derived from the standard deviation within each frequency bin, though this estimate reflects only the variance within a given realization. Alternative error estimates based on PSDs computed from different time segments would likely result in comparable uncertainties on the fitted slopes.

At fixed plasma $\beta$, the PSD slope systematically decreases with increasing deformation angle; for example, at $\beta = 0.001$, $\alpha$ changes from $1.25 \pm 0.07$ ($30^\circ$) to $0.76 \pm 0.06$ ($60^\circ$), representing a substantial systematic change across models. Similarly, at fixed inclination, variations with $\beta$ are significant: at $30^\circ$, $\alpha$ increases from $1.25 \pm 0.07$ ($\beta = 0.001$) to $1.59 \pm 0.06$ ($\beta = 0.007$), indicating a consistent trend with plasma $\beta$ that exceeds the level of stochastic variability within individual simulations.

The impact of numerical resolution is also assessed: comparing the $\beta = 0.001$, $30^\circ$ run with a higher-resolution simulation, the slopes differ ($\alpha = 1.25 \pm 0.07$ vs. $0.92 \pm 0.06$), highlighting a sensitivity of the absolute PSD slope to numerical resolution and realization, although the overall red-noise character remains unchanged. This indicates that while the precise slope values are not fully converged, the qualitative variability properties are robust.

The combination of magnetic field deformation and asymmetric outflows plays a role in shaping the PSD. At low deformation angles, polar regions dominated by wind outflows introduce transient, localized variability that enhances high-frequency power, thereby reducing the PSD slope. This effect weakens at higher deformation angles, consistent with a reduction in polar-dominated variability.

%\section*{Appendix C}
%\subsection*{\MakeUppercase{C. Horizon Magnetic Flux Asymmetry}}
\section{Horizon Magnetic Flux Asymmetry}
\label{app_D}

To investigate whether the black hole retains memory of the initial magnetic field configuration, we compute the normalized horizon flux difference,
\begin{equation}
\frac{\Delta \Phi}{\Phi_{\rm tot}} = \frac{\Phi_{\rm north} - \Phi_{\rm south}}{\Phi_{\rm north} + \Phi_{\rm south}},
\end{equation}
where $\Phi_{\rm north}$ and $\Phi_{\rm south}$ are the fluxes threading the northern and southern hemispheres of the horizon, respectively. A value of $\Delta \Phi/\Phi_{\rm tot} = 0$ corresponds to a perfectly symmetric split-monopole configuration.

\begin{figure}
\centering
\includegraphics[width=0.48\textwidth]{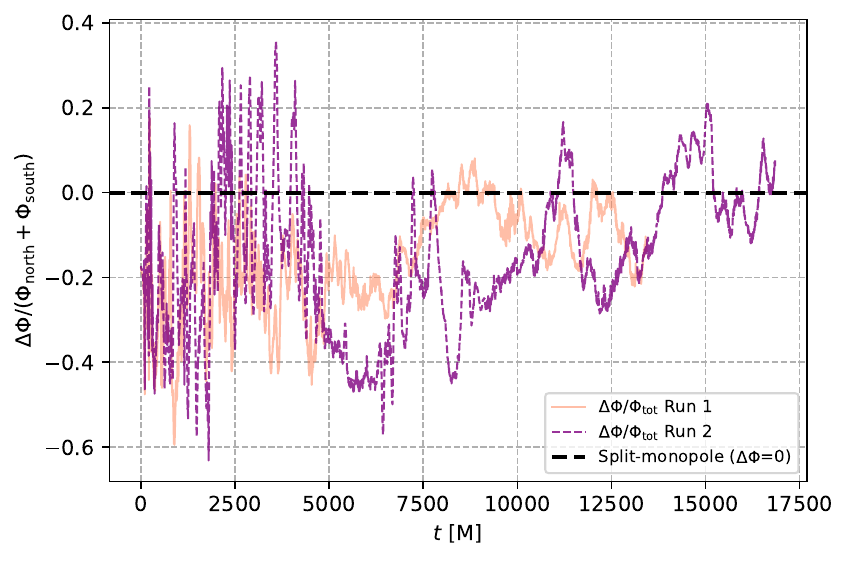}
\caption{Time evolution of the normalized horizon flux difference $\Delta \Phi/\Phi_{\rm tot}$ for two simulations with an initial $30^\circ$ magnetic field deformation. Run 1 (lower resolution, $1024 \times 512$) is shown in orange, and Run 2 (higher resolution, $2048 \times 1024$) is shown in purple. The dashed black line corresponds to a perfectly symmetric split-monopole configuration ($\Delta \Phi = 0$). The system does not fully relax to a split-monopole, likely due to initial deformations.}
\label{fig:delta_phi_vs_t}
\end{figure}

Figure~\ref{fig:delta_phi_vs_t} shows the normalized horizon flux difference $\Delta \Phi/\Phi_{\rm tot}$ for two 2D simulations with $\beta = 0.001$ and an initial magnetic field deformation of $30^\circ$: a lower-resolution run ($1024 \times 512$, Run~1) and a higher-resolution run ($2048 \times 1024$, Run~2). Over most of the simulated duration, the horizon-threading magnetic flux does not settle into a perfectly symmetric split-monopole configuration. Instead, a persistent north–south asymmetry is maintained, reflecting the continuous influence of the accretion flow on the magnetic field topology. While non-accreting, oblique magnetospheres are expected to relax to a split-monopole configuration within $\sim 300\,r_g/c$ \citep{2024ApJ...968L..10S, 2025ApJ...988L..33S}, our simulations include a dynamically accreting flow, where continuous turbulence, magnetic reconnection, and flux advection actively perturb the field geometry, preventing the system from quickly reaching the split-monopole state. We find that the system begins to approach a more symmetric configuration only at much later times ($t \gtrsim 14000\,t_g$), suggesting that the accretion flow plays a key role in preserving memory of the initial field deformation over extended periods. Fully three-dimensional simulations will be required to determine whether this delayed relaxation persists when non-axisymmetric modes are allowed.

%\section*{Appendix D}
%\subsection*{\MakeUppercase{D. Resolution Comparison}}

\bibliography{reference}
\bibliographystyle{aasjournalv7}

\end{document}